\documentclass[pra,twocolumn,amssymb]{revtex4}

\usepackage{graphicx}
\usepackage{amssymb,amsmath} 
\usepackage{appendix}

\begin{document}

\title{Scaling of ground state fidelity in the thermodynamic limit: XY model and beyond}

\author{Marek M. Rams$^{1,2}$ and Bogdan Damski$^1$} 
\affiliation{$^1$Los Alamos National Laboratory, 
Theoretical Division, MS B213, 
Los Alamos, New Mexico, 87545, USA \\
$^2$Institute of Physics, Jagiellonian University, Reymonta 4, PL-30059 Krak\'ow, Poland} 
\begin{abstract}
We study ground state fidelity defined as the overlap between  two 
ground states of the  same quantum system obtained for slightly different values of the 
parameters of its Hamiltonian. We focus on the thermodynamic regime of the XY model and the neighborhood of its critical points. We describe in detail cases when fidelity is dominated by  the universal contribution reflecting quantum criticality of the phase transition. We show that  proximity to the multicritical point leads to anomalous scaling of fidelity. 
We also discuss   fidelity in a regime characterized by pronounced oscillations resulting from the change of either the system size or the parameters of the Hamiltonian. Moreover, we show when fidelity is dominated by non-universal contributions, 
study fidelity in the extended Ising model, 
and  illustrate how our results provide additional insight into dynamics of quantum phase transitions.
Special attention is put to studies of fidelity from the momentum space perspective. All our main results are obtained analytically. They are in excellent agreement with numerics.

\end{abstract}
\maketitle

\section{Introduction}
\label{Sec_Intro}
Quantum phase transitions (QPTs) happen when dramatic changes in 
the ground state properties of a quantum system can be induced by 
a tiny  variation of an external parameter  \cite{sachdev}. 
This system-specific parameter can be a magnetic field in spin systems
\cite{Coldea2010,sadler2006etal}, intensity of a laser beam in cold atom simulators of 
Hubbard-like models \cite{Hubbard_exp}, dopant concentration 
in high-Tc superconductors \cite{Lee2006}, etc. 

Traditional condensed matter approaches to QPT focus on 
identification of the order parameter and the pattern of symmetry breaking
as well as on studies of two point correlation functions  and the excitation gap \cite{sachdev}.  
At the critical point of the second order QPT   
the correlation length diverges while the gap in the excitation spectrum 
vanishes. This is typically described by power-law 
singularities. Correlation length $\xi$  diverges as  $|\lambda-\lambda_c|^{-\nu}$, while the excitation gap 
closes as $|\lambda-\lambda_c|^{z\nu}$, where $\lambda$ is the external field driving the 
transition, $\lambda_c$ marks the quantum critical point, and $z$ and 
$\nu$ are the critical exponents associated with the universality class of the system.
The exponents are universal in the sense that they do not dependent on the microscopic details 
of the system. 

A somewhat different ways of looking at  QPTs have recently emerged from quantum information 
community. 
One of them is based on studies of quantum entanglement of
spins/atoms/etc. undergoing a QPT \cite{Osterloh2002}. 
Another is based on studies of ground state fidelity \cite{Zanardi2006} (see \cite{GuReview} for a recent review).
Finally, an extreme approach emerged  where the Hamiltonian is designed to have 
a particular many-body state as its ground state \cite{Wolf2006}. We will
study ground state fidelity below.

More precisely, ground state fidelity, or simply fidelity, is  defined here as 
\begin{equation}
{\cal F}(\lambda,\delta) = |\langle \lambda-\delta|\lambda+\delta\rangle|,
\label{def_F}
\end{equation}
where $|\lambda\rangle$ is a ground state wave function of a many-body Hamiltonian
$\hat H(\lambda)$ describing the system exposed to an external field $\lambda$, while 
$\delta$ is  a parameter difference.
It provides the 
most basic probe into the dramatic change of the wave function 
around the  critical point 
and 
has the following  properties
\begin{equation}
0\le {\cal F}(\lambda,\delta) \le 1, \label{Fbounds}  \ \ \ {\cal F}(\lambda,\delta) = {\cal F}(\lambda,-\delta). 
\end{equation}
The recent surge in studies of fidelity follows observation  that 
quantum criticality promotes its decay \cite{Zanardi2006}. This is easily justified 
as  ground states change rapidly near the critical point 
to reflect  singularities of a  QPT. Therefore, one expects that ${\cal
F}(\lambda,\delta)$ has  a minimum at the critical point.

The drop in fidelity encodes not only the position of the critical point, but also 
universal information about the transition given by the critical exponent
$\nu$. This has been worked out independently in
the ``small system'' \cite{Venuti2007,Schwandt2009,ABQ2010,BarankovArXiv2009,deGrandi2010a,deGrandi2010b,PolkovnikovArXiv2010} 
and in the thermodynamic  \cite{MMR2010} limits.
Broadly speaking the former corresponds to $\delta\to0$ at fixed system size $N$, while the latter corresponds to $N\to\infty$ at
small, fixed $\delta$. 

In the ``small system'' limit we can Taylor expand fidelity in $\delta$ \cite{Zanardi2006,You2007,GuReview} 
\begin{equation}
{\cal F}(\lambda,\delta) \simeq 1-\delta^2 \chi_F(\lambda)/2,
\label{fid_sus}
\end{equation}
where $\chi_F$ stands for the so-called fidelity susceptibility. 
The linear term in $\delta$ disappears in the above expansion due 
 to the normalization condition of the ground states whose overlap 
is taken (see e.g. \cite{GuReview}). This can be also seen  from 
 (\ref{Fbounds}):  a linear in $\delta$ term could make ${\cal F}>1$ 
or simply fidelity is symmetric with respect to the $\delta\to-\delta$ transformation.
Universal information can be extracted from fidelity susceptibility  near the critical 
point through the scaling:  $\chi_F(\lambda_c) \sim N^{2/d \nu}$, where $d$ stands for 
system dimensionality. Alternatively, one may look at $\chi_F$
away from the critical point and study $\chi_F(\lambda) \sim N|\lambda-\lambda_c|^{d\nu-2}$.
More generally, the scaling of fidelity susceptibility is linked to the  scaling dimension of the most relevant perturbation \cite{Venuti2007}.

In the thermodynamic limit one obtains \cite{MMR2010}:
 \begin{equation}
\ln{\cal F}(\lambda,\delta) \simeq  -N|\delta|^{d\nu} A\left(\frac{\lambda-\lambda_c}{|\delta|} \right),
\label{near_F}
\end{equation}
where $A$ is a scaling function. 
We expect the above scaling result to hold when fidelity per lattice site ($-\ln{\cal
F}/N$) is well defined in the thermodynamic limit, 
 physics around the critical point is described by one characteristic 
scale of length (correlation length) given by the critical exponent $\nu$,
and $d\nu<2$ so that non-universal (system-specific) corrections 
to (\ref{near_F}) are subleading \cite{deGrandi2010a,deGrandi2010b,PolkovnikovArXiv2010}. 

\noindent
Expression (\ref{near_F}) can be simplified  away from
the critical point when $|\delta| \ll |\lambda-\lambda_c| \ll 1$:
\begin{equation}
\ln{\cal F}(\lambda,\delta) \sim -N\delta^2 |\lambda-\lambda_c|^{d\nu -2}.
\label{away_F}
\end{equation}
Unlike (\ref{fid_sus}), (\ref{near_F}) is nonanalytic in $\delta$ near the
critical point and the Taylor expansion of fidelity in $\delta$ is
inapplicable here. 
This singularity arises from anomalies associated with the 
QPT in the thermodynamic limit. All these results have been obtained in the zero
temperature limit that we adopt in this work. Even though a consistent theory of fidelity in finite temperatures
is still missing, several results have already been obtained
\cite{Sirker2010,QuanFidTemp2007, FernandoFidTemp2009}.

This article is organized as follows. Sec. \ref{Sec_intro2} describes  
theoretical approaches used to study fidelity and 
motivation behind this research.
In Sec. \ref{Sec_XYmodel} we focus on fidelity in the XY model showing numerous analytical results. 
Sec. \ref{Sec_MPS} discusses  fidelity of the extended Ising model \cite{Wolf2006,Cozzini2007,Zhou2008}.
In Sec. \ref{instant_quench} we illustrate some connection between fidelity and 
dynamics of a nonequilibrium quench.
Our findings are briefly summarized in Sec. \ref{Sec_con}.


\section{Basics of fidelity}
\label{Sec_intro2}

To start, we mention  that the term fidelity is used in various
contexts in quantum physics to describe the similarity between two  quantum
states. Therefore, it is important to
remember that we use it  only to refer to an overlap between two ground states
of the same physical system calculated for different values of its external parameters.

One of the seminal results on fidelity was obtained by Anderson
decades ago \cite{Anderson1967}.  He showed on a particular model that
fidelity
disappears in the thermodynamic limit. Similar 
behavior has been later found in other models and was labeled as the Anderson
orthogonality
catastrophe. More importantly, orthogonality catastrophe was shown to 
play a role in  numerous  condensed matter
systems (see e.g. \cite{Wiegmann2006} and references therein). 

While being an equilibrium quantity, fidelity shows up in nonequilibrium
dynamics of  quantum systems. In particular, one encounters it 
in the context of nonequilibrium QPTs (see \cite{DziarmagaReview,polkovnikov_review2010} for reviews on dynamics of QPTs).
For example, the scaling of density of excited quasiparticles following a quench 
can be derived using it  \cite{deGrandi2010a,deGrandi2010b,PolkovnikovArXiv2010}.  Moreover, 
an envelope of nonequilibrium coherence oscillations in a central spin problem 
encodes fidelity as well \cite{BD2009}.

In the context of equilibrium QPTs, fidelity has been proposed as an efficient theoretical 
probe of quantum criticality (the so-called fidelity approach). 
To appreciate its simplicity, we compare the fidelity approach 
to the traditional method of studies of quantum criticality based on analysis 
of the asymptotic decay of the two point correlation
functions. First, both approaches require the same input: the ground state wave
functions. Second, they provide the same information:  
the location of the critical point and the universal critical exponent $\nu$.
Third, the conventional  approach is based on the assumption that the 
two point correlations between spins/atoms/etc. decay semi-exponentially
away from the critical point and  in a power law manner at the critical point \cite{Vojta2003}.
The transition  from 
semi-exponential to the power-law decay can be tedious to study  even with the recent
tensor network techniques (see e.g. \cite{ZakrzewskiBHqcp}). 
Therefore, the fidelity
approach is arguably a simpler alternative at least as far as numerical
calculations are concerned. 

On the analytical side, compact and accurate results for fidelity are scarce. 
Typically, models are not exactly solvable and so analytical results are out of reach for them.
In exactly solvable systems situation is far from trivial as well. Indeed, in systems like the Ising model one is left with a large product of
analytically known factors that have only recently been cast into a simple 
expression: see \cite{MMR2010} and Sec. \ref{Sec_XYmodel}. 
The situation is much different in  some
systems where ground states
can be exactly expressed through finite rank 
Matrix Product States: see  \cite{Wolf2006,Cozzini2007,Zhou2008}
and Sec. \ref{Sec_MPS}. The Hamiltonians of these systems, however,  are
``engineered'' to have a pre-determined ground state \cite{Wolf2006}. These
difficulties motivate various approximate approaches.

First, powerful numerical techniques have been deployed including 
tensor networks  \cite{Zhou2008_Vidal} and Quantum Monte Carlo
\cite{Schwandt2009,ABQ2010} simulations. These approaches provide 
crucial insight into models that are not exactly solvable (e.g. \cite{Schwandt2009}), 
and shall be especially useful for studies of systems
with unknown order parameters, critical points and critical exponents. 

Second, the fidelity susceptibility approach based on the Taylor expansion
(\ref{fid_sus}) has been used. Simplification here comes from factoring out the parameter difference
$\delta$ and focusing on fidelity susceptibility that
depends on the external field only. Despite these simplifications, it is typically still a mixed analytical and 
numerical technique. This approach is limited to 
studies of fidelity between very similar ground states whose overlap is  close to
unity. In particular, this rules out description of Anderson orthogonality catastrophe within  this framework.

Third, the fidelity per lattice site approach has been proposed 
\cite{Zhou2008,Zhou2008_Ising,Zhou2008_Vidal}.
It is in a sense ``orthogonal'' to the fidelity susceptibility approach
as (i) it targets overlap between any two ground states: any $\delta$ is
 considered; (ii) it focuses on the large system limit, i.e.,
 the opposite of what the fidelity susceptibility approach does; and (iii) 
 fidelity typically departs significantly from unity here.
 This approach  has been explored mostly numerically so far.

Our studies assume two most plausible features 
of the above two approaches \cite{MMR2010}. 
First, the thermodynamic limit from the fidelity per lattice site approach which allows us to study 
fidelity in systems well approximating the critical (infinite) ones. 
Second, small $\delta$ as in 
fidelity susceptibility approach allowing for derivation of universal scaling results 
such as (\ref{near_F}) and  (\ref{away_F}). Moreover, combination of the two assumptions allows for derivation of  compact analytical  results for 
fidelity in some exactly solvable models: the task presumably impossible otherwise. 

\section{XY model}
\label{Sec_XYmodel}
In this section we are going to study fidelity of the XY model 
\begin{equation}
\hat H = 
-\sum_{n=1}^N\left( \frac{1+\gamma}{2} \sigma^x_n\sigma^x_{n+1} +\frac{1-\gamma}{2} \sigma^y_n\sigma^y_{n+1} +  g \sigma^z_n \right),
\label{XYH}
\end{equation}

\noindent where we assume periodic boundary conditions $\sigma_{N+1} = \sigma_{1}$. 
This model is exactly solvable via the Jordan-Wigner transformation translating it into a 
free fermionic system
\cite{sachdev,Lieb1961,BarouchMcCoy1971,Bunder1999PRB}.

Above $g$ is the external magnetic field acting along the $z$ direction and  $\gamma$ is the anisotropy of 
spin-spin interactions on the $xy$ plane.
The critical exponent $\nu$ for Ising-like critical points, $g=\pm1,\gamma\neq0$  and $-1< g <1,\gamma=0$ is  
\begin{equation}
\nu=1.
\label{Ising_uni}
\end{equation}

\noindent 
The specific case of a multicritical point located at $g=\pm1$ and $\gamma=0$ needs special attention \cite{damle1996}.
It has been  proposed that there are 
two divergent characteristic length scales  that have to be taken into account around it: one 
characterized by the scaling exponent $\nu=1/2$ and the other by $\nu=1$ \cite{Mukherjee2011}.

\begin{figure}
\includegraphics[width=\columnwidth,clip=true]{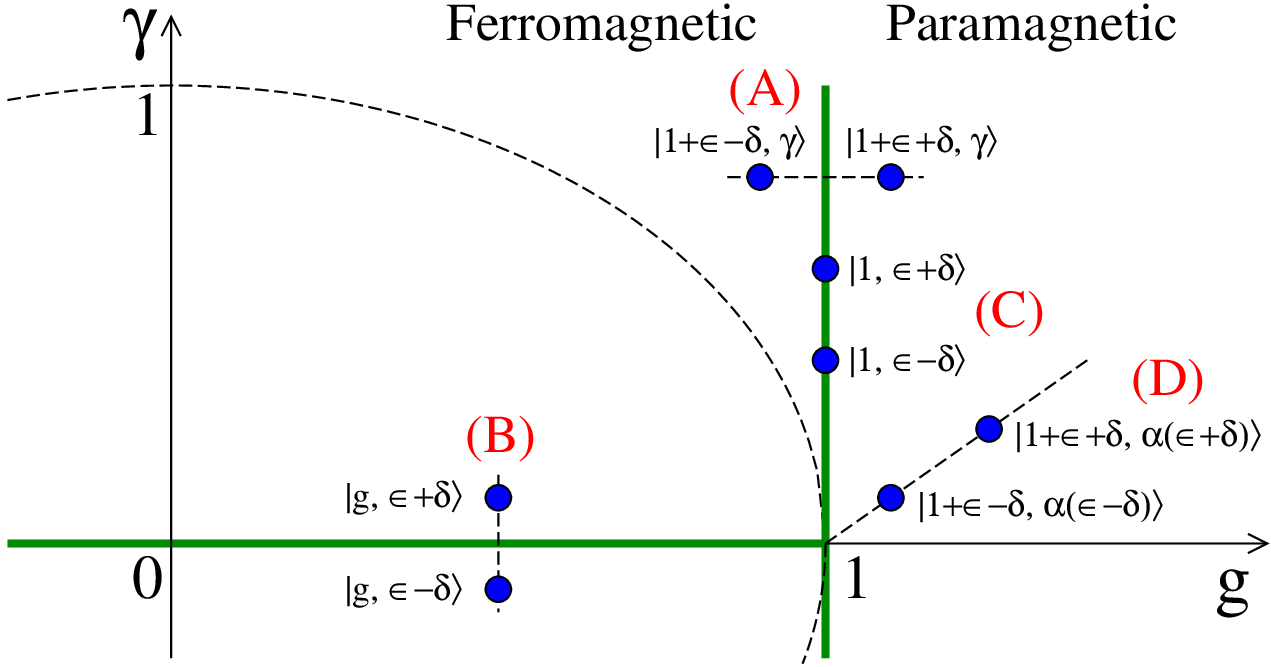}
\caption{(Color online). Phase diagram of the XY model (\ref{XYH}). Critical points lie on green (thick) lines.
(A) to (D) show different characteristic lines along which we calculate fidelity. 
The dashed ``circle"  separates two regions of the ferromagnetic phase differing by the structure of the excitation gap. 
Minimum of the energy gap is reached for $k_c=0$  outside the ``circle"  ($k_c=\pi$ for $g<0$) and for  
$k_c=\arccos[g/(1-\gamma^2)]$ inside the ``circle" (see e.g. \cite{Bunder1999PRB}).}
\label{fig1}
\end{figure}

 We will study here 
\begin{equation}
{\cal F} = |\langle g_1,\gamma_1|g_2,\gamma_2\rangle|,
\label{FXY}
\end{equation}
where $|g,\gamma\rangle$ is the ground state of (\ref{XYH}).
As depicted in Fig. \ref{fig1}, we choose $(g_i,\gamma_i)$ to lie on  straight lines near critical 
points/lines.

The analytic expression for fidelity in the XY model was given 
in \cite{Zanardi2006}: 
${\cal F} = \prod_{k>0}\left|\cos\left(\theta^1_k/2-\theta^2_k/2\right)\right|$, where 
$\tan \theta^i_k =\gamma_i \sin k/\left(g_i-\cos k\right)$ and momenta $k$ are given by (\ref{Fprod_k}).
We have reworked it to the form which proves to be
convenient for subsequent analytical studies
\begin{eqnarray}
&{\cal F}&=\prod_{k>0} f_k,  \label{Fprod} \\ 
&k& = (2n+1)\pi/N, \  n = 0,1,\ldots,N/2-1,\label{Fprod_k}\\ 
&f_k&= \sqrt{\frac{1}{2}+\frac{1}{2} \frac{p_k}{\sqrt{p_k^2+q_k^2}}},\label{Fprod_fk}\\
&p_k& =  (g_1-\cos k)(g_2-\cos k)+\gamma_1 \gamma_2 \sin^2 k, \label{Fprod_pk}\\
&q_k& =  [\gamma_2 (g_1-\cos k)-\gamma_1 (g_2-\cos k)] \sin k, \label{Fprod_qk}
\end{eqnarray}
which is valid for any $g_1$, $g_2$, $\gamma_1$ and $\gamma_2$.
To be more precise, we assume even $N$ and follow notation from
\cite{JacekPRL2005} during diagonalization of (\ref{XYH}). The ground state lays then in a subspace with even number of quasiparticles which leads to (\ref{Fprod_k}). In that subspace the XY Hamiltonian (\ref{XYH})  is diagonalized to the
form $\hat H=\sum_{k>0 } \epsilon_k \left( \gamma_k^\dagger \gamma_k + \gamma_{-k}^\dagger \gamma_{-k}  -1 \right)$, where
$\gamma_k$ are fermionic annihilation operators  and the energy gap is
\begin{equation}
\epsilon_k = 2\sqrt{(g-\cos k)^2+\gamma^2 \sin^2 k}.
\label{energy_gap}
\end{equation}

In the leading order fidelity (\ref{Fprod})  can be additionally simplified by replacing 
the product over momentum modes by an integral
\begin{equation}
\ln{\cal F}  ~=~ \sum_{k>0} \ln f_k ~\simeq~ \frac{N}{2 \pi} \int_{0}^{\pi} dk  \ln f_k,
\label{Fintegral}
\end{equation}
which is allowed in the thermodynamic limit as all the integrals that we study 
are  convergent everywhere (even at the critical points). We note here, however, that in some cases $\ln f_k$ has a logarithmic singularity adding a subleading term to the transformation from summation into integration (\ref{Fintegral}). 
This will be discussed in details in Secs. \ref{Sec_Ising} and \ref{Sec_XX}.

\subsection{Across the g=1 critical line}
\label{Sec_Ising}
In this paragraph we follow the path A from Fig. \ref{fig1}, i.e., we substitute 
\begin{equation}
g_{1,2}=1+\epsilon\pm\delta, \ \gamma_1=\gamma_2=\gamma, \ \epsilon = c |\delta|
\label{Apath}
\end{equation}
into (\ref{FXY}) and furthermore assume that $0<\gamma\le 1$. 
The relevant critical point  is at $g_c=1$ with the critical exponent given by (\ref{Ising_uni}).
The relevant correlation length was calculated in \cite{BarouchMcCoy1971} and reads 
\begin{equation}
\xi \sim \left| \frac{1}{\ln|(g-\sqrt{g^2-1+\gamma^2})/(1-\gamma)|} \right|.
\label{XYC}
\end{equation}

\noindent As long as  $\gamma^2 \gg |g-1|$, it can be approximated near the critical $g_c=1$ by 
\begin{equation}
\xi \sim \frac{\gamma}{|g-1|}.
\label{XY_coh}
\end{equation}
To use this expression for the correlation length and to find the leading, universal behavior of fidelity we keep $|\delta|,|\epsilon| \ll \gamma^2$ in this section.
We note that it implies that we stay outside of the 
``circle" in Fig. \ref{fig1}. 
In this region the minimum of the energy gap is reached for the smallest value of momentum $k$ (\ref{Fprod_k}). 
This condition also keeps us away from the multicritical point at $\gamma=0$ and $g=1$,
which will be investigated in Secs. \ref{Sec_NON} and \ref{Sec_Lif}.

The results of this section complement our previous studies from \cite{MMR2010} where we have considered a special case of the Ising model: $\gamma=1$.
Here -- besides extending our results to $\gamma\neq1$ -- we discuss the problem in momentum space, show details of our calculations and estimate errors of our
approximations.

To begin the discussion, 
we expect  the system to crossover from the ``small system'' limit to  the thermodynamic limit when  \cite{MMR2010}
\begin{equation}
\min[\xi(\lambda+\delta), \xi(\lambda-\delta)] \sim L,
\label{tlimit}
\end{equation}
where $\lambda$ denotes a point on the $(g,\gamma)$ plane, while $\delta$
is its displacement. 
The latter is in general a two-component vector.
Above $\xi(\lambda)$ is the correlation length in a system exposed to the external field $\lambda$ and $L$ is the linear size of the system: $L^d=N$. 
Fidelity approaches the thermodynamic limit result when  
$$\min[\xi(\lambda+\delta), \xi(\lambda-\delta)] \ll L.$$
On the other hand, when 
$$\min[\xi(\lambda+\delta), \xi(\lambda-\delta)] \gg L$$  
the ``small system" limit is reached  and the behavior of fidelity is dominated by finite size effects. We use the quotation marks to highlight that we still consider $L\gg1$ here as QPTs shall not be studied in systems made of a few spins/atoms/etc.

\begin{figure}
\includegraphics[width=\columnwidth,clip=true]{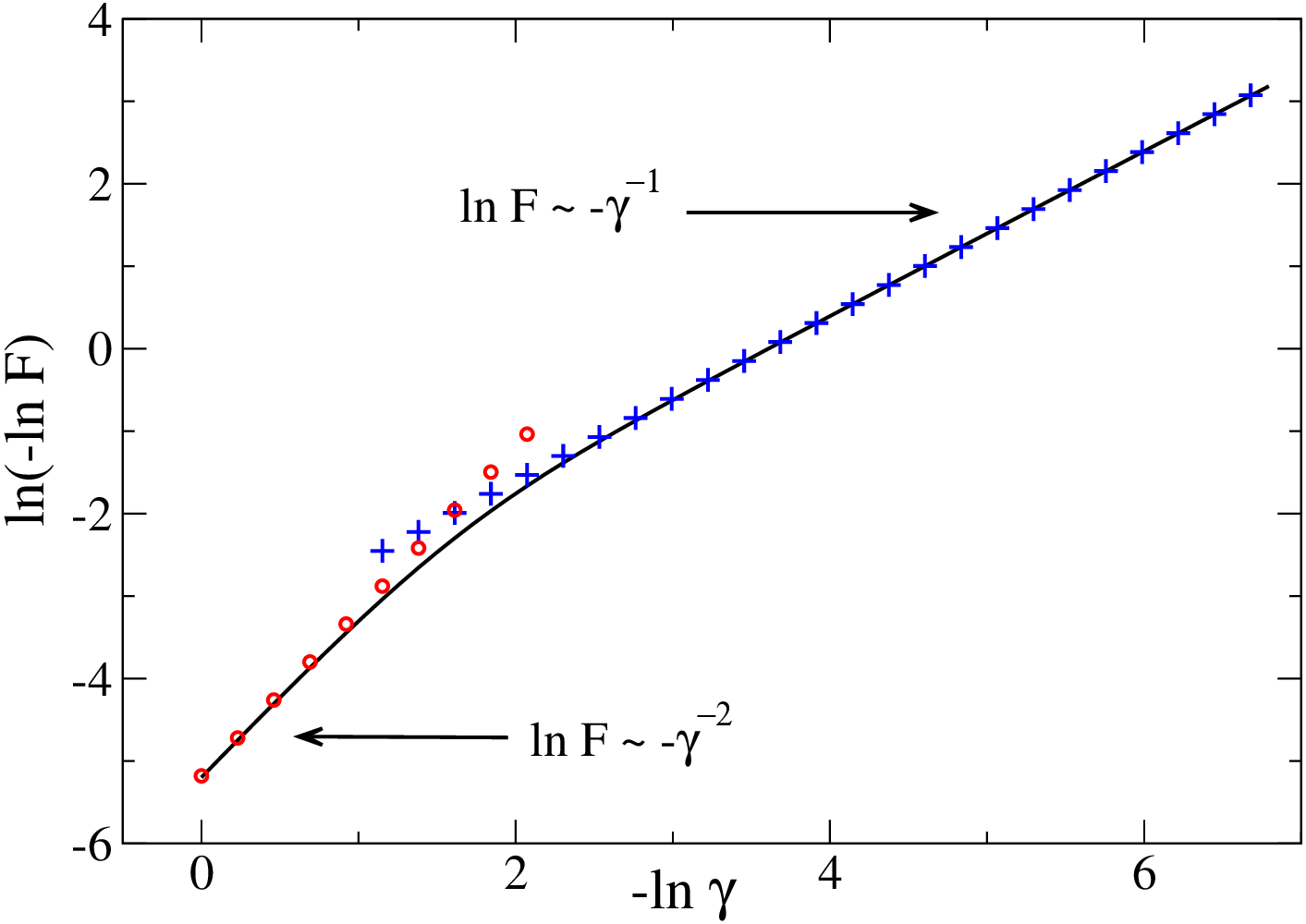}
\caption{(Color online). Transition from the ``small system'' limit (left part
of the plot) to the thermodynamic limit (right part of the plot) resulting from variation of the anisotropy parameter 
$\gamma$. Solid black line is a numerical result showing ${\cal F}=|\langle g_c -2 \delta,\gamma | g_c, \gamma \rangle| $, 
red dots depict the ``small system'' 
prediction (\ref{XY_sus}), blue crosses represent the thermodynamic result
(\ref{z}). The figure is  prepared for $\delta=3\times10^{-7}$ and  $N=10^6$.
}
\label{fig2}
\end{figure}

We notice that the 
correlation length in (\ref{XY_coh}) has a prefactor which is linear in $\gamma$. By changing it we will illustrate the crossover (\ref{tlimit}). 
To that end, we fix $N$, $\delta$ and $\epsilon=-|\delta|$ ($c=-1$) so that one of the states is exactly at the critical point,  i.e.  ${\cal F} = \left|\langle g_c-2\delta,\gamma | g_c,\gamma\rangle\right|$. 
The outcome of such a calculation is depicted in Fig. \ref{fig2}.
Two distinct regimes are visible there.  
In the left part of the plot, where $\gamma\approx1$, we observe $\ln{\cal F} \sim -\gamma^{-2}$. In the right
part of the plot, where $\gamma\ll1$, we find $\ln{\cal F}\sim -\gamma^{-1}$. These two regimes correspond to the ``small system''
limit and the thermodynamic limit, respectively.

\begin{figure}
\includegraphics[width=\columnwidth,clip=true]{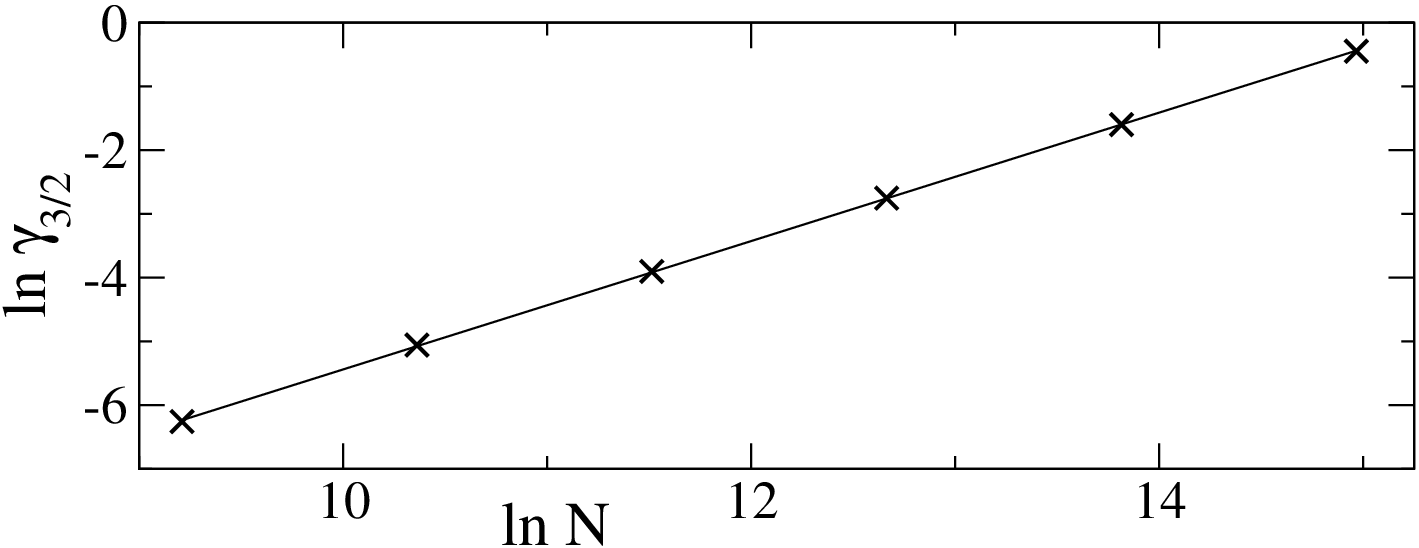}
\caption{Numerical study of the crossover condition (\ref{Ising_thermo}): see
text for details.
Crosses come from numerics, while straight line is a linear fit:
$\ln\gamma_{3/2} = -15.51\pm0.05 + (1.007\pm0.004)\ln N$ \cite{Fit}. The fit shows that 
the crossover takes place when $\gamma \sim N$ as predicted by
(\ref{Ising_thermo}).
We used here $\delta=3\times10^{-7}$ as in Fig. \ref{fig2} and extracted
$\gamma_{3/2}$ from ${\cal F}=|\langle g_c -2 \delta,\gamma | g_c, \gamma \rangle|$.
}
\label{fig3}
\end{figure}

Substituting (\ref{Apath}) and (\ref{XY_coh}) (with fixed $c$) 
into (\ref{tlimit}) we obtain the condition for the crossover as
\begin{equation}
\frac{N|\delta|}{\gamma}\sim 1.
\label{Ising_thermo}
\end{equation}
The validity of the above condition is numerically
confirmed in Fig. \ref{fig3}. This is done in the following way. 
As  $-\ln\gamma$ is increased in Fig. \ref{fig2},  
the slope of $\ln(-\ln{\cal F})$  
changes smoothly from $2$ (corresponding to $\ln{\cal F}\sim -\gamma^{-2}$) to $1$
(corresponding to $\ln{\cal F}\sim -\gamma^{-1}$). The crossover region
between the two limits is centered around $\gamma=\gamma_{3/2}$ where the local slope equals $3/2$.
By repeating the calculation from Fig. \ref{fig2} for various system sizes
$N$, but the same $\delta$, we have numerically obtained $\gamma_{3/2}(N)$.
A power-law fit described in Fig. \ref{fig3} confirms 
that $\gamma_{3/2}\sim N$.  Complementary  
analysis has been done in \cite{MMR2010} where $\gamma=1$ and $\delta$-dependence of 
(\ref{Ising_thermo}) has been verified. Based on these two calculations 
 we conclude that the crossover condition  
(\ref{Ising_thermo}) holds near the Ising critical point. In fact, it is 
in an excellent agreement with our numerical simulations.

After describing the crossover between the two regimes observed in 
Fig. \ref{fig2}, we can explain in detail the scaling  observed 
on both ``ends'' of the plot. We start with the ``small system'' limit.
A simple analytical calculation based on the expansion (\ref{fid_sus}) -- i.e. the
fidelity susceptibility approach -- shows that near the critical point  
\begin{equation}
{\cal F}\simeq 1 - \frac{\delta^2N^2}{16\gamma^2} \Rightarrow \ln{\cal F}\sim \gamma^{-2}
\label{XY_sus}
\end{equation}
because fidelity is very close to unity here. 
As illustrated in Fig. \ref{fig2}, this result agrees well with numerics. 
For completeness, we mention that away from the critical point, $|\epsilon|\gg|\delta|$, expansion (\ref{fid_sus}) yields
\begin{equation}
{\cal F}\simeq 1 - \frac{ \delta^2 N}{16\gamma |\epsilon|},
\label{XY_sus2}
\end{equation}
in the leading order in $|\epsilon| \ll \gamma^2$. 

Explanation of the thermodynamic limit result requires more involved calculations.
Substituting  (\ref{Apath}) into (\ref{Fprod_pk}-\ref{Fprod_qk}) we obtain
\begin{eqnarray}
p_k&=&(1+\epsilon-\cos k)^2-\delta^2+\gamma^2 \sin^2 k, \label{pk1} \\
q_k&=&2\delta \gamma \sin k. \label{qk1}
\end{eqnarray}
To calculate $\ln{\cal F}=\ln |\langle g_c+\epsilon+\delta,\gamma| g_c+\epsilon-\delta,\gamma\rangle |$ in the leading universal order in $\delta$ we 
notice that the integral (\ref{Fintegral}) is dominated by the contribution from small momenta $k$. This allows us 
 to approximate (\ref{pk1},\ref{qk1}) by $p_k\approx k^2 \gamma^2 + \epsilon^2-\delta^2$ and 
 $q_k \approx 2 \delta \gamma k$.  Thus, we get  
\begin{equation}
\ln  f_k \approx \frac{1}{2}\ln\left(\frac{1}{2} + \frac{k^2 \gamma^2 +\epsilon^2-\delta^2 }{2\sqrt{(k^2 \gamma^2 +\epsilon^2-\delta^2)^2+(2\delta\gamma k)^2}}\right).
\label{tildefk_Ising}
\end{equation}

\noindent Next we put (\ref{tildefk_Ising})  into (\ref{Fintegral}), 
change the integration variable $k$ there to $l~|\delta| / \gamma$, and send
the new upper integration limit,  $\pi \gamma / |\delta|$,  to $\infty$. 
After all these approximations we obtain 
\begin{equation}
\ln {\cal F} \simeq 
\frac{N |\delta|}{4 \pi  \gamma} \int_0^\infty dl \ln \left( \frac{1}{2}+\frac{l^2+c^2-1}{2\sqrt{(l^2+c^2-1)^2+4 l^2}}  \right).
\label{fid_ising_app}
\end{equation}
\noindent The above integral can be calculated analytically,
\begin{equation}
\frac{\ln{\cal F}}{N} \simeq - |\delta| A(c)/\gamma,
\label{z}
\end{equation}
where the scaling function
\begin{equation}
A(c)= \left \{
\begin{array}{cc} 
\begin{split}
& \frac{1}{4}+ \frac{|c| K(c_1)}{2\pi}+\frac{(|c|-1) {\rm Im} E(c_2)}{4\pi}& {\rm for}~ |c|<1,  \\
& \frac{|c|}{4} - \frac{|c| K(c_1)}{2\pi}-\frac{(|c|-1) {\rm Im} E(c_2)}{4\pi}& {\rm for}~ |c|\ge1,
\end{split}
\end{array}
\right. 
\label{Ac}
\end{equation}
and  
\begin{equation}
c_1 = -4 |c|/(|c|-1)^2, \ \ c_2=(|c|+1)^2/(|c|-1)^2.
\label{c1c2}
\end{equation}
The complete elliptic integrals of the first and second kind are defined as 
\small
\begin{equation}
K(x)=\int_0^{\pi/2} \frac{d\phi}{\sqrt{1-x \sin^2 \phi}},  E(x)=\int_0^{\pi/2} d\phi \sqrt{1-x \sin^2 \phi},
\label{KEeliptic}
\end{equation}
\normalsize
respectively. This result is quite interesting. 

First, solution (\ref{z}) compares well to numerics, which is illustrated
in Fig. \ref{fig2} for $c=-1$. Not only the relation $\ln{\cal F}\sim -\gamma^{-1}$ is
reproduced, but the whole expression  fits numerics
 well. We mention in passing that the same good agreement was obtained for
$\gamma=1$ and different $\delta$'s  in \cite{MMR2010}, which confirmed the 
general scaling prediction (\ref{near_F}). Notice that not the anisotropy $\gamma$ but the shift in the magnetic field $\delta$ is the {\it relevant} perturbation here.

Second, a singularity of derivative of fidelity can be obtained from 
(\ref{z}): $\frac{d}{dc}\ln{\cal F}|_{c\to1^\pm}$
is logarithmically divergent (see \cite{MMR2010} for $\gamma=1$ case).
To make a more transparent connection to the former studies of Zhou and 
collaborators who called such
singularities as the pinch points \cite{Zhou2008,Zhou2008_Vidal,Zhou2008_Ising}, we consider  
$\frac{\partial}{\partial g_2} \tilde d(g_1,g_2)$, where  the scaling parameter is defined as 
\begin{equation}
\tilde d(g_1,g_2) = -\lim_{N\rightarrow\infty} \frac{\ln {\cal F}}{N}.
\label{tilded}
\end{equation}

\noindent This can be analytically calculated 
from (\ref{z}) to be 
\begin{equation}
\frac{\partial}{\partial g_2}\tilde d(g_1,g_2) =
\left.\frac{dA}{dc}\right|_{c=\epsilon/\delta} \frac{\epsilon+\delta}{2\delta \gamma} -\frac{1}{2 \gamma}
A\left(\frac{\epsilon}{\delta}\right),
\label{derivative}
\end{equation}
where the relations between $g_{1,2}$, $\varepsilon$ and $\delta$ are
specified in (\ref{Apath}). Above we assume $g_1>g_2$ i.e. $\delta>0$.
The explicit expression for (\ref{derivative}) is quite involved and so we do
not show it. It predicts a logarithmic singularity when $g_2 \rightarrow 1$ with prefactors which depend on $g_1 \neq 1$.
More precisely, this singularity arises in the limit of $N\to\infty$, already assumed in (\ref{derivative}), 
and is rounded off for finite systems: see Fig. \ref{fig4}. As will be
shown below, it  has an interesting interpretation in
momentum space.

\begin{figure}
\includegraphics[width=\columnwidth,clip=true]{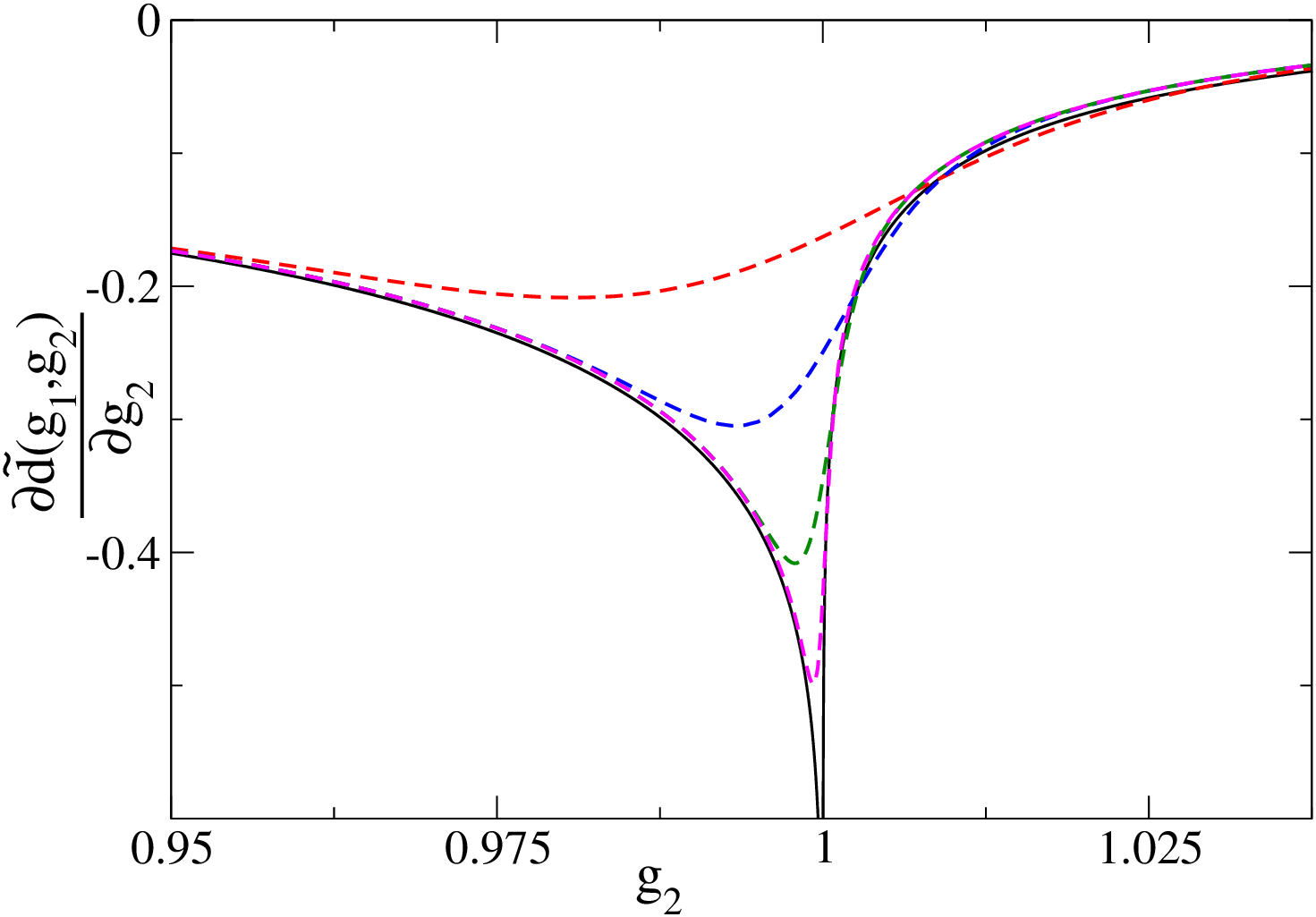}
\caption{(Color online).  Derivative of the scaling parameter:
$\frac{\partial}{\partial g_2} \tilde d(g_1,g_2)$. Solid black line is
the analytical result (\ref{derivative}). Dashed lines from top to bottom 
correspond to numerical results for $N=100,300,1000,3000$, respectively.
The field $g_1=1.1$ and $\gamma=1$.
}
\label{fig4}
\end{figure}

Third, solution (\ref{z}) can be simplified away from the critical point. 
Straightforward expansion done for  $|c| \gg 1$ (but still $|\epsilon| = |c\delta| \ll \gamma^2$), 
gives $A(c)\simeq 1/16 |c|$ and so 
$$
{\cal  F}~\simeq~\exp(-N\delta^2/16\gamma|\epsilon|)
$$
in agreement with (\ref{away_F}). 
When the argument of the exponent is not small, this is 
a new result. 
Otherwise, it coincides with expression (\ref{XY_sus2}), from the fidelity susceptibility approach (\ref{fid_sus}).

\begin{figure}
\includegraphics[width=\columnwidth,clip=true]{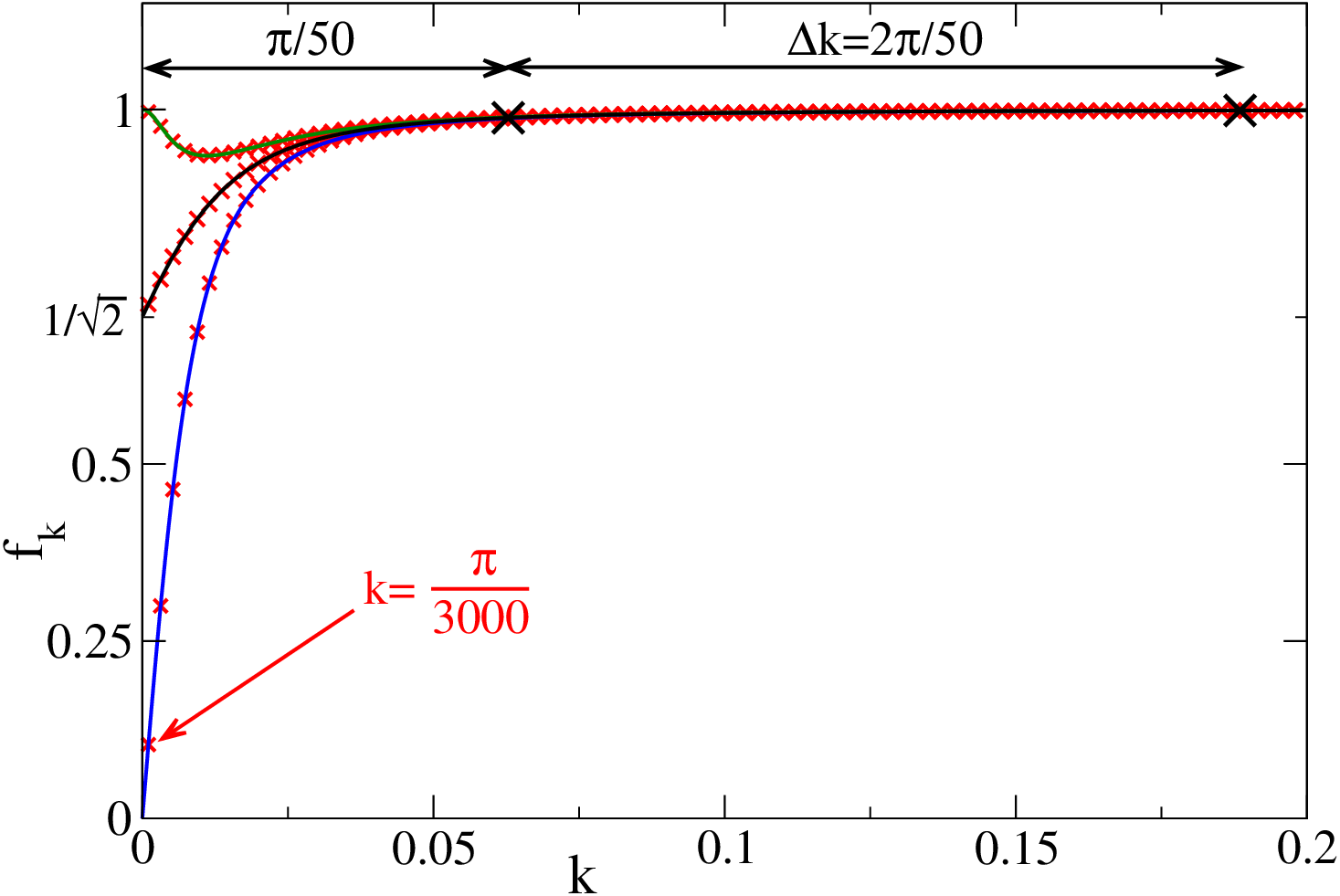}
\caption{(Color online). 
Thermodynamic vs. ``small system'' limit near  the Ising 
critical point. 
Solid lines correspond to $f_k$ calculated for $\gamma=1$, $\delta=0.01$,  and -- from bottom to top  -- 
$c$ equals 0, 1 and 1.5,  respectively. 
Small red crosses mark $f_k$'s  contributing to fidelity calculated for $N=3000$. 
Large black crosses show the same, but 
when the system size is $N=50$. 
For $N=3000$ the thermodynamic limit is reached. 
$N=50$ represents the ``small system'' limit.  
Notice that in the ``small system'' limit the resolution in momentum space is insufficient to account for abrupt changes of $f_k$ taking place
at small momenta.}
\label{fig5}
\end{figure}

Complementary insight into fidelity can be obtained in momentum space where we focus on  $f_k$.
 It is presented in Fig. \ref{fig5} where we show prototypical behavior of $f_k$ for the Ising model when:
(i) the ground states entering fidelity are calculated on the opposite sides of the critical
point $|c|<1$;  (ii) one of them is obtained  at the critical point $|c|=1$; 
(iii) they are both obtained on  the same side of the critical point $|c|>1$. 

 As illustrated in Fig. \ref{fig5},  if $\Delta k = 2 \pi/N \gg |\delta|/\gamma$  
 only $f_k\approx 1$ contribute to fidelity, and we end up in the ``small system'' limit.  The system is too small to monitor changes of $f_k$ between zero and unity. Still, we are able to observe universal finite size effects 
 \cite{Schwandt2009,ABQ2010,BarankovArXiv2009,deGrandi2010a,deGrandi2010b,PolkovnikovArXiv2010}
 in (\ref{XY_sus}).
 
\noindent In the opposite limit of  $\Delta k \ll |\delta|/\gamma$, we are able to approach the exact ($N\rightarrow \infty$) 
thermodynamic limit  and the product (\ref{Fprod})  is well approximated by the integral (\ref{Fintegral}) [but see the discussion below as well]. 
Now momenta $k$ are dense enough to monitor leading changes in $f_k$ between zero and unity.

\begin{figure}
\includegraphics[width=\columnwidth,clip=true]{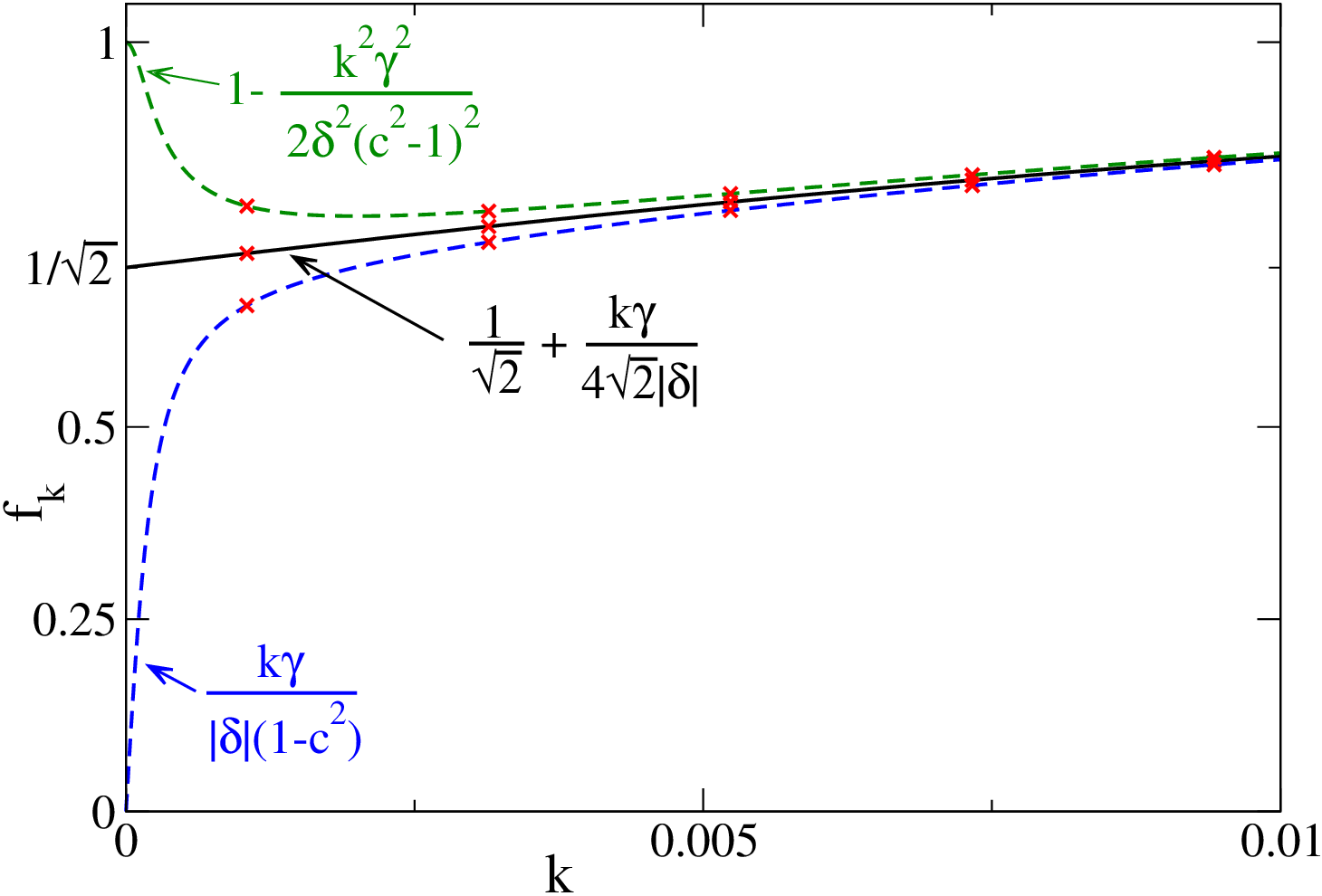}
\caption{(Color online). Discontinuous change of $f_k$ 
around $c=1$. The curves from bottom to top correspond to 
$c$ equal to $0.98$, $1$ and $1.02$, respectively. 
Red crosses, similarly as in Fig. \ref{fig5},
mark $f_k$'s  contributing to fidelity calculated for $N=3000$. 
The parameter shift $\delta=0.01$.
Lowest order Taylor expansions of $f_k$ are listed (but not plotted) for all three curves.
}
\label{fig6}
\end{figure}

Still, to see sharp nonanalyticity (i.e. the pinch point) 
in fidelity when one of the states passes through the critical point we need to have $N\rightarrow \infty$ as demonstrated in Fig. \ref{fig4}.  
It is caused by discontinuity of $f_k$ at zero
momentum when $|c|=1$. Indeed, from (\ref{Fprod_fk},\ref{pk1},\ref{qk1}) it is easy to calculate the limits
\begin{equation}
\begin{array}{lr}
\lim_{k\to k_c} f_k = 0 & {\rm for}~  |c|<1, \\
\lim_{k\to k_c} f_k = \frac{1}{\sqrt{2}} & {\rm for}~ |c|=1, \\
\lim_{k\to k_c} f_k = 1 & {\rm for}~ |c|>1,
\end{array}
\label{disco}
\end{equation}
where $k_c=0$ here. 
This is presented in Figs. \ref{fig5} and \ref{fig6}. In the latter, it is shown that the jump of $f_k$ is roughly happening on the momentum scale given by the inverse of the larger correlation length scaling as 
$|\delta (|c|-1)|/\gamma$, 
which is divergent at $|c| \rightarrow 1$ (notice that we have two correlation lengths here for the two ground states 
entering fidelity).
It results in rounding off of the derivative of the scaling parameter $\tilde d= -\ln {\cal F} /N$ around the critical point for 
finite $N$ due to insufficient sampling of $f_k$ near $k_c=0$.
We also notice that the approximation
(\ref{tildefk_Ising}) that we make to obtain (\ref{z}) correctly capture this discontinuity.

Finally, we can discuss errors resulting from our approximations. 
On the one hand, approximating exact expression for $f_k$ (\ref{Fprod_fk},\ref{pk1},\ref{qk1}) with (\ref{tildefk_Ising}) in the integral (\ref{Fintegral}) leads to ${\cal O}(\delta^2/\gamma^3)$  error in (\ref{z}). 
This calculation is quite technical and has been deferred to the  Appendix. 
On the other hand, there are errors connected with estimation of the product (\ref{Fprod}) by the integral (\ref{Fintegral}). 
Main contribution here comes from the logarithmic divergence of 
$\ln f_k$ at $k=0$ when 
$$
|c|<1.
$$ 
It results in a subleading shift which has to be added to the right-hand side of (\ref{Fintegral}). 
This shift  saturates to $\ln 2/2$ when the size of the system $N$ is large enough 
to well sample the logarithmic singularity.  
It happens when the system size 
$N$ is much larger than the larger of the two correlation lengths proportional to $\gamma/|(1-|c|) \delta|$.
It can be seen, e.g., when we expand $f_k$  to the lowest order in $k$ around $k_c=0$,
\begin{equation}
f_k \approx  \frac{\gamma k}{|\delta| (1-c^2)}, 
\label{fk_expansion}
\end{equation}
for $|c|<1$. Using this result we obtain
\begin{equation}
\sum_{0<k<k_{\rm cutoff}} \ln (\alpha k) - \frac{N}{2 \pi} \int_0^{k_{\rm cutoff}} dk
\ln (\alpha k)  \overset{N\rightarrow \infty}{=} \frac{\ln 2}{2}. 
\label{log_shift}
\end{equation}

\noindent Above, $\alpha = \gamma/|\delta|(1-c^2)$, $k$'s are given by (\ref{Fprod_k}) and 
$k_{\rm cutoff}$  restricts summation/integration to small momenta for which 
the expansion (\ref{fk_expansion}) is meaningful. 
We get the equality in (\ref{log_shift}) in the limit of $k_{\rm cutoff}/\Delta k \to \infty$  
when the logarithmic singularity is sampled densely. This can be easily obtained using the 
Stirling formula.

\begin{figure}
\includegraphics[width=\columnwidth,clip=true]{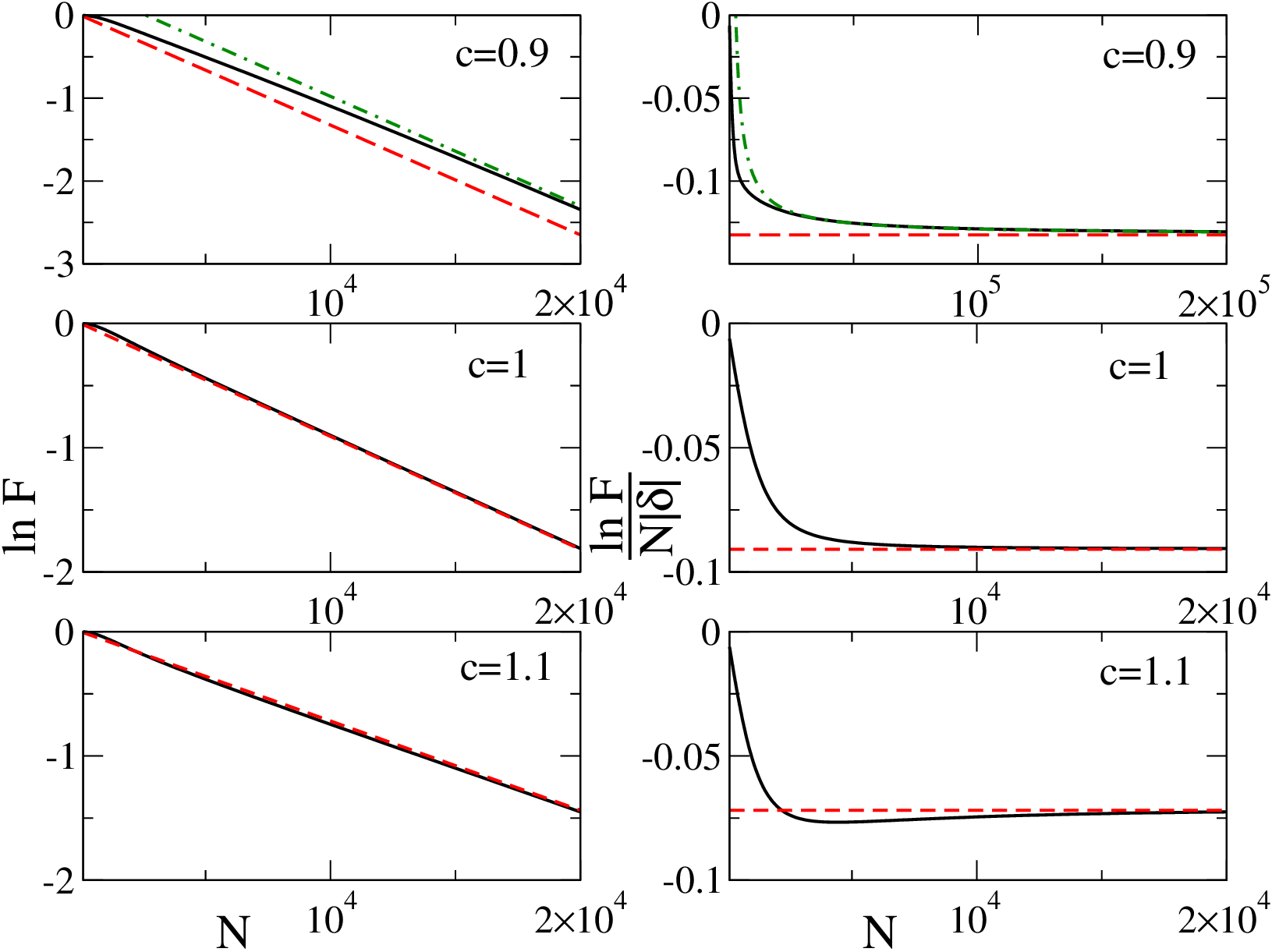}
\caption{(Color online).  
Panels in the left and right column illustrate 
$\ln {\cal F}$ and $\ln{\cal F}/N|\delta|$ as a function of $N$, respectively.
From top to bottom rows show results for  $c$ equal to $0.9$, $1$ and $1.1$.
Black solid lines show numerics. Red dashed lines present $-A(c) N |\delta|$ in the left 
column and $-A(c)$ in right column.
Green dashed-dotted line in the top row shows $-A(c)N|\delta|+\ln 2/2$ in the
left panel and $-A(c)+\ln2/2N|\delta|$ in the right one.
See discussion around  (\ref{lnFcfinal}) and (\ref{Fcfinal}) for more
information about the logarithmic correction. 
The parameter shift $\delta=10^{-3}$ and $\gamma=1$. 
}
\label{fig7}
\end{figure}

Including the above discussed errors/corrections into (\ref{z}) we obtain 
\begin{equation}
\frac{\ln\cal F}{N} = \left \{
\begin{array}{ll} 
 -|\delta| A(c)/\gamma  +  \ln2/2N + {\cal O}(\delta^2/\gamma^3)& \ {\rm for}~|c|<1,  \\
 -|\delta| A(c)/\gamma +  {\cal O}(\delta^2/\gamma^3)& \ {\rm for}~|c|\ge1,
\end{array}
\right.
\label{lnFcfinal}
\end{equation}
which is equivalent to 
\begin{equation}
{\cal F} = \left \{
\begin{array}{cc} 
\begin{split}
& \sqrt{2} \exp\left[-N|\delta| A(c)/\gamma +  N{\cal O}(\delta^2/\gamma^3)\right]&{\rm for}~|c|<1,  \\
&  \exp\left[-N |\delta| A(c)/\gamma +  N{\cal O}(\delta^2/\gamma^3)\right]&{\rm for}~|c|\ge1.
\end{split}
\end{array}
\right. 
\label{Fcfinal}
\end{equation}
Note that the difference between $|c|<1$ and $|c|\ge1$ cases comes from the
lack of a logarithmic singularity in $\ln f_k$ in the latter. As illustrated
in Fig. \ref{fig7}, 
numerics compares very well to  (\ref{lnFcfinal}) and (\ref{Fcfinal}). 
The prefactor $\sqrt{2}$ is seen in numerical simulations 
when $N \gg  \gamma/|(1-|c|) \delta|$. In derivation of (\ref{lnFcfinal}) and
(\ref{Fcfinal}) we have neglected other errors coming from changing the product (\ref{Fprod}) into the integral 
(\ref{Fintegral}). 
Those corrections are present around $c=\pm 1$ but 
are subleading and  disappear in the limit of $N\rightarrow \infty$ (e.g. they
smooth out the pinch point nonanalyticity).

Finally, we  would like to stress that a subleading correction to the argument of the exponent
in (\ref{lnFcfinal}), the $\ln2/2N$ term, increases fidelity by the $\sqrt{2}$ factor (\ref{Fcfinal}). 
Its influence can be neglected when we study the scaling parameter, $-\ln{\cal F}/N$ 
from (\ref{lnFcfinal}), instead of fidelity. This is presented in the right column of Fig. \ref{fig7}.

\subsection{Across $\gamma$=0 critical line}
\label{Sec_XX}
In this paragraph we follow the path B from Fig. \ref{fig1}, i.e., 
we substitute into (\ref{FXY}) 
\begin{equation}
\gamma_{1,2}=\epsilon\pm\delta, \ g_{1,2}=g\in(-1,1),  \ \epsilon = c |\delta|.
\label{Bpath}
\end{equation}
Two remarks are in order now. First, we shift 
$\gamma$ instead of $g$ here, but use the same symbols, $\delta$ and
$\epsilon$, for the shift as in Sec. \ref{Sec_Ising}. Therefore,
some care is required to avoid confusion. Second, critical 
exponents $z$ and $\nu$ are the same as for  the Ising
universality class studied in Sec. \ref{Sec_Ising}. Other critical
exponents, however, differ between  the two transitions \cite{Bunder1999PRB}. 

Expanding (\ref{XYC}) in small $\gamma$ relevant for our calculations, 
the correlation length reads
$$
\xi\sim \gamma^{-1},
$$
when $1\gg \gamma < \sqrt{1-g^2}$. This can be used to predict the crossover condition from (\ref{tlimit}).
We focus here on large enough $N$ to study the thermodynamic limit. 
Furthermore, we restrict our calculations to $|\epsilon|,|\delta|\ll \sqrt{1-g^2}$ 
in order to make the problem analytically tractable.
 
\begin{figure}
\includegraphics[width=\columnwidth,clip=true]{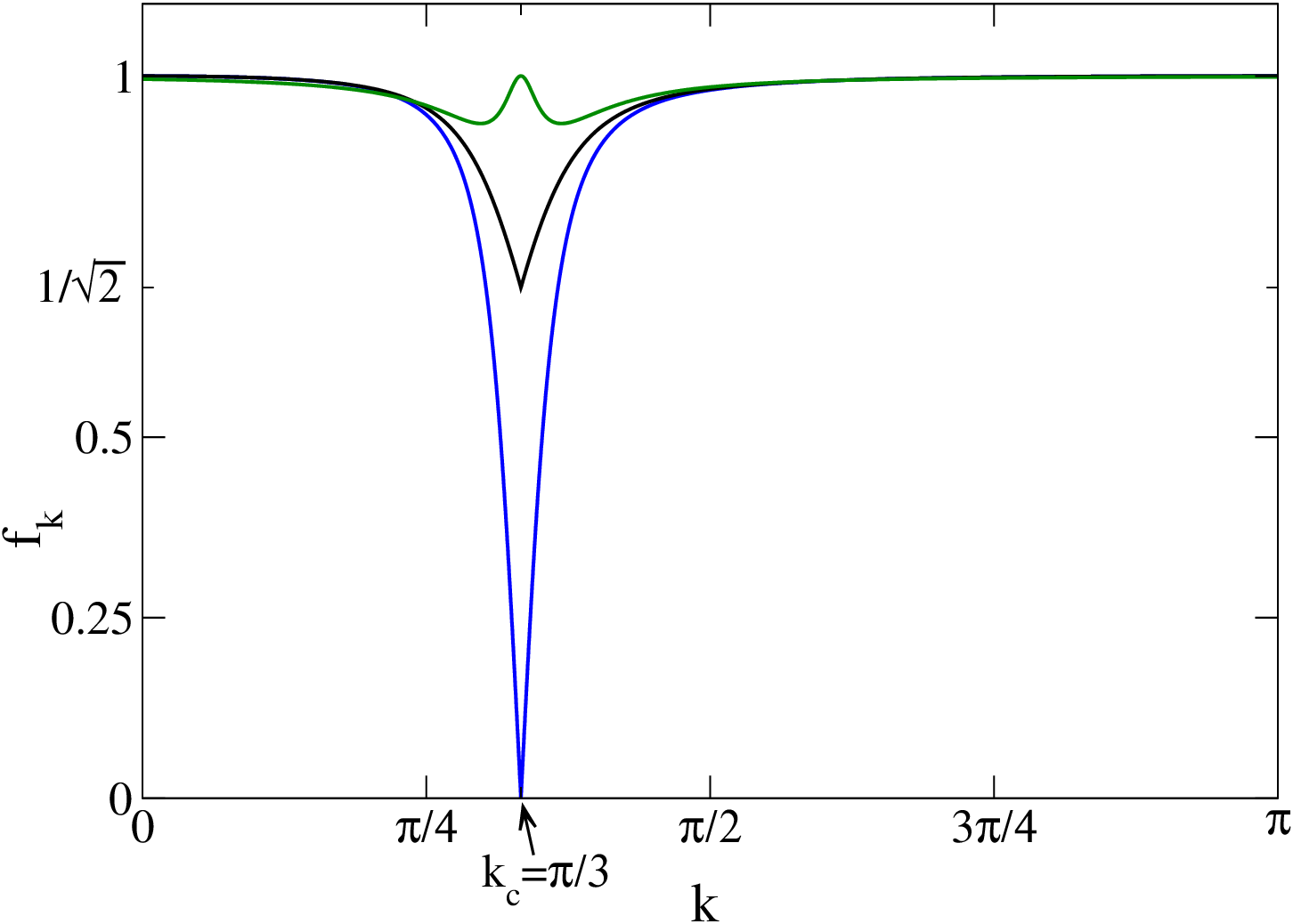}
\caption{(Color online). 
Plot of $f_k$ 
for the transition across the $\gamma=0$ line (\ref{Fprod_fk}).
We choose  $g=0.5$ and $\delta=0.1$ here. 
The curves from bottom to top correspond to $c$ equal to $0$, $1$, and $1.5$, respectively.  
Dominating contribution to fidelity is centered around 
$k_c = \arccos(g)=\pi/3$. 
}
\label{fig8}
\end{figure}

The crucial difference with respect to the calculations from Sec.
\ref{Sec_Ising} is that at the critical points -- lying on the $\gamma=0$ and $g\in(-1,1)$ line --
the gap in the excitation spectrum (\ref{energy_gap}) closes for momentum 
 \begin{equation}
 k_c=\arccos(g). 
 \label{An_kc}
 \end{equation}
As can be expected, fidelity in the thermodynamic limit is dominated by contributions 
coming from momenta centered around $k_c$ (Fig. \ref{fig8}). 

Putting (\ref{Bpath}) into (\ref{Fprod_pk},\ref{Fprod_qk}) one gets the following exact expressions
\begin{eqnarray}
p_k &=& (g-\cos k)^2 + \left(\epsilon^2 -\delta^2 \right) \sin^2 k, \label{pk2} \\
q_k &=& 2 \delta (g-\cos k) \sin k.\label{qk2}
\end{eqnarray}
We expand the above around dominant $k_c$ to get $p_k\approx (1-g^2) \left((k-k_c)^2 +\epsilon^2 - \delta^2 \right)$ and $q_k\approx (1-g^2)  2 \delta (k-k_c)$,  where we have utilized the condition $|\delta|, |\epsilon| \ll \sqrt{1-g^2}$. Thus, we get
{\small
\begin{equation}
\ln f_k\approx
\frac{1}{2}\ln\left(\frac{1}{2} + \frac{(k-k_c)^2 +\epsilon^2-\delta^2 }{2\sqrt{((k-k_c)^2 +\epsilon^2-\delta^2)^2+4\delta^2 (k-k_c)^2}}  \right).
\label{szlag}
\end{equation}
}

\begin{figure}
\includegraphics[width=\columnwidth,clip=true]{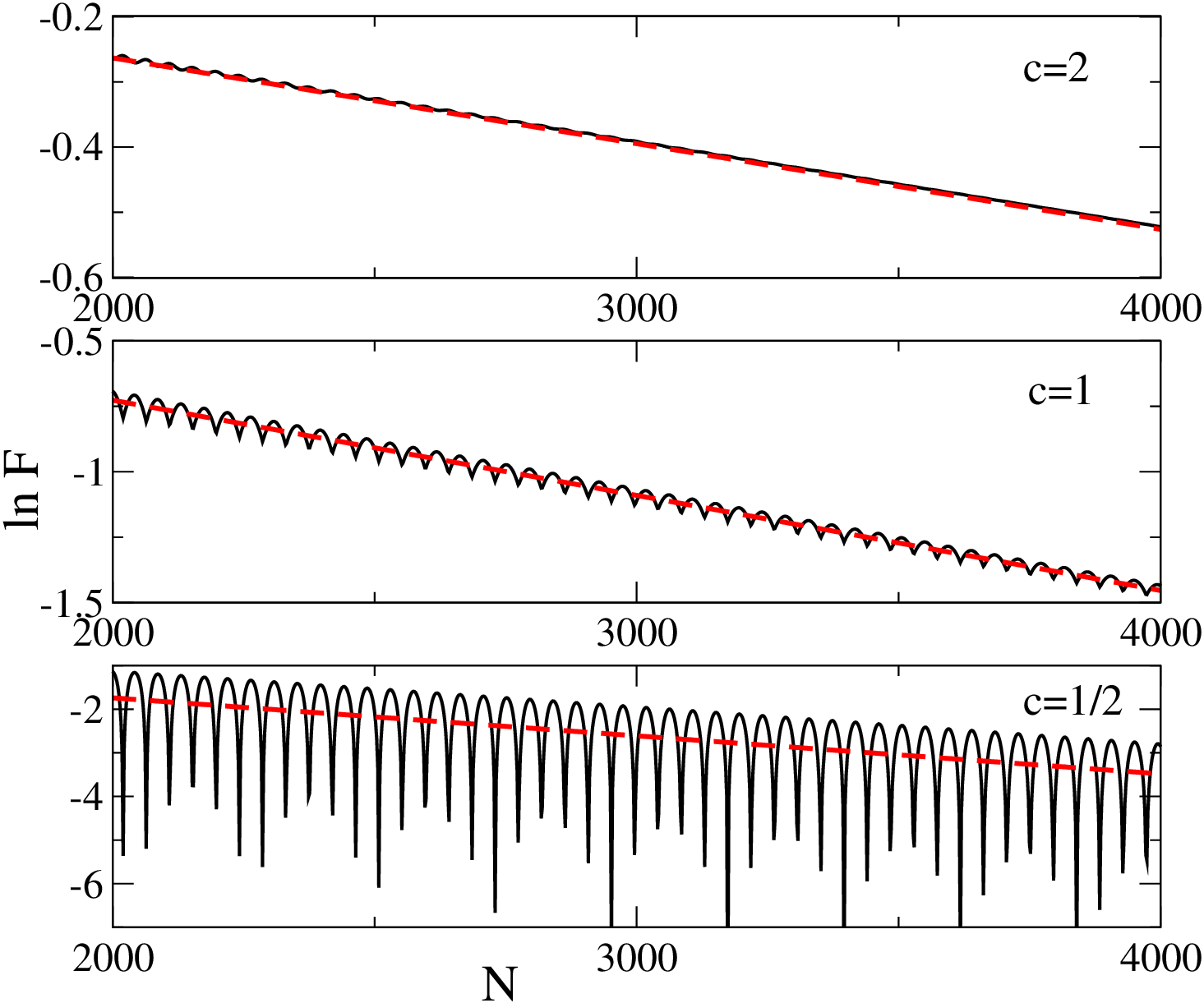}
\caption{(Color online). 
Comparison between the smooth part of fidelity  (\ref{fid_isingXX}) 
-- given by the  red dashed line -- and the exact numerical result (solid
black line). In the large $N$-limit and for $|c|>1$ oscillations of fidelity are absent,
which is presented here for $c=2$ (top panel). At $c=1$ some oscillations of fidelity 
are visible for
not-too-large-systems (middle panel), but they gradually disappear for large $N$.
For $|c|<1$ pronounced oscillations of fidelity surviving in the thermodynamic 
exist: bottom panel prepared for  $c=1/2$. The plots are made  for 
$\delta = 0.002$ and $g=0.99$.  
}
\label{fig9}
\end{figure}

\noindent We integrate it over $k$  (\ref{Fintegral}), 
parameterize $\epsilon$ as $c|\delta|$ (note that $\epsilon$ is now the shift 
of the anisotropy parameter), 
change the variable $k$ in (\ref{Fintegral}) to $k_c +l|\delta|$, 
and send the new integration limits, $-k_c/|\delta|$ and $(\pi-k_c)/|\delta|$,
to $\mp\infty$ respectively.  

\noindent After all these approximations we arrive at 
\small
\begin{equation}
\ln {\cal F}_{\rm smooth} \simeq 
\frac{N |\delta|}{4 \pi} \int_{-\infty}^\infty dl \ln \left( \frac{1}{2}+\frac{l^2+c^2-1}{2\sqrt{(l^2+c^2-1)^2+4 l^2}}  \right).
\label{fid_XX_app}
\end{equation}
\normalsize
We have introduced above the symbol ${\cal F}_{\rm smooth}$ because
it turns out that for $|c|<1$, i.e. when the two states used to calculate fidelity are obtained on opposite sides
of the critical point, there will be oscillatory corrections to fidelity (Fig. \ref{fig9}). 
Before describing them, we 
analyze ${\cal F}_{\rm smooth}$. 

\noindent Except for the integration range, the above integral is exactly the same as the one for the Ising model in 
(\ref{fid_ising_app}). Therefore, we obtain
\begin{equation}
\frac{\ln{\cal F}_{\rm smooth}}{N} = -|\delta|  2A(c) +  {\cal O}\left(\delta^2 \right),
\label{fid_isingXX}
\end{equation}
where $A(c)$  is defined in (\ref{Ac}). 
We estimated numerically the error between the exact integral (\ref{Fintegral}) and the approximated one 
(\ref{fid_XX_app}), what is  presented in the Appendix.
Note that (\ref{fid_isingXX}) quantifies decay rate of fidelity with system
size even when the oscillations are present (Fig. \ref{fig9}).

As pointed out in e.g. \cite{Bunder1999PRB}, 
the anisotropic transition has the same critical behavior as a pair of decoupled Ising chains.
This can be seen also 
through fidelity: there is a prefactor of two in (\ref{fid_isingXX}) absent in (\ref{z}) from the previous section.
Looking from the momentum space perspective, the factor of two comes from different location of the 
momentum $k_c$.  
For the Ising model considered in Sec. \ref{Sec_Ising}, $k_c=0$ and only $k> k_c$ contribute to fidelity (\ref{Fprod}). 
Here  $k_c\neq0$ and so both $k>k_c$ and $k<k_c$ add up
to fidelity doubling the result [additivity of this
effect can be seen from (\ref{Fintegral})]. 

We mention also that the pinch point singularity is predicted by 
(\ref{fid_isingXX}). It is again visible in momentum space through
discontinuity of $f_k$ and described by  (\ref{disco}): notice that now
$k_c\neq0$ (\ref{An_kc}). This is illustrated in Fig. \ref{fig8}.

\begin{figure}
\includegraphics[width=\columnwidth,clip=true]{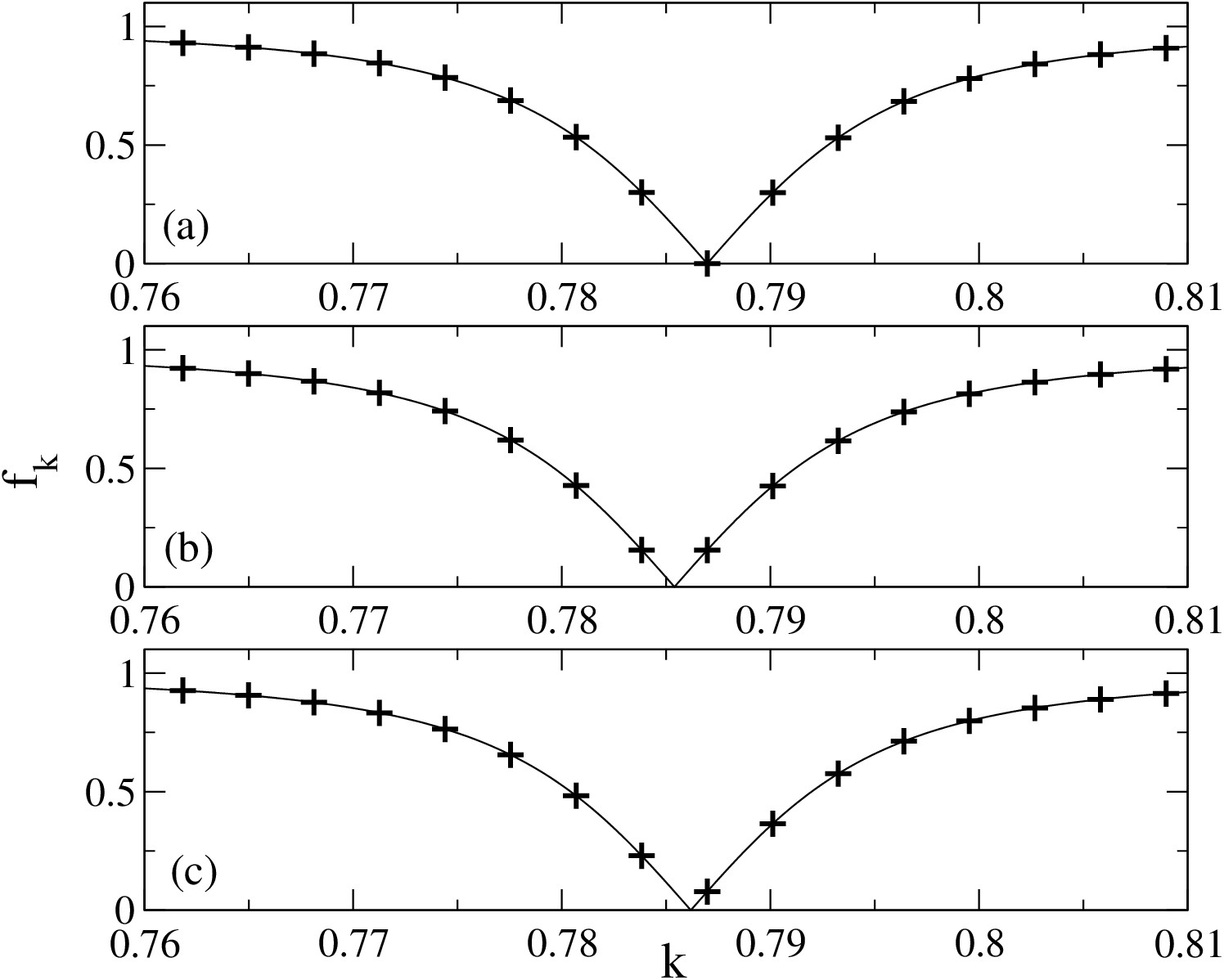}
\caption{Plot of $f_k$ (\ref{Fprod_fk}) for the transition across the $\gamma=0$ 
line for $\delta=0.01$, $c=0$ and several values of $g$
such that (a) $k_c=\pi\frac{1002}{4000}$, (b) $k_c=\pi/4$, and 
(c) $k_c=\pi\frac{1001}{4000}$ [note that $k_c$ depends on g (\ref{An_kc})].
Pluses mark discretization in momentum space for the size of the system $N=2000$. 
We see a systematic shift of $f_k$'s contributing to fidelity as we slightly change $g$. This 
results in oscillations of fidelity from the lowest panel of Fig. \ref{fig9}. 
Parameter $\phi$ from (\ref{An_phi}) equals $1$, $0$ and $1/2$ for panels (a), (b), and (c), respectively.
}
\label{fig10}
\end{figure}

Now we are ready to describe the oscillatory corrections to (\ref{fid_isingXX}) depicted 
in Fig. \ref{fig9}.
These effects, visible for 
$$
|c|<1,
$$
are washed away by the above continuous  approximations relying on the limit of $N\to\infty$  
that overlooks discretization of  momenta $k$. 
The reason behind the oscillatory behavior is easily identified by looking at 
 Fig. \ref{fig10}. Indeed, fidelity is a product of $f_k$'s (\ref{Fprod}) and 
 as such it is sensitive to the $f_k$ factors approaching zero. For
 $|c|<1$,  $f_k$'s stay close to zero when $k\approx k_c$ 
 (as shown in Fig. \ref{fig8} this is not the case for $|c|\ge1$). The arrangement of $k$'s
 around $k_c$ -- fixed by discretization (\ref{Fprod_k}) --  determines fidelity oscillations. 

Qualitatively, fidelity reaches its minimum at zero when one of $k$'s  equals 
exactly $k_c$. This happens when $k_c= (2 n+1)\pi/N$ for some integer $n\in[0,N/2-1]$.
Then $f_{k=k_c} = 0$ and a  single momentum mode takes fidelity down to
zero even for finite $N$. This case is illustrated in Fig. \ref{fig10}a. On the
other hand, when $k_c$ lays symmetrically between two discretized $k$'s, 
Fig. \ref{fig10}b, we can expect that fidelity is near its maximum.
All other ``orientations'' of discretized momenta $k$ with respect to
$k_c$, including the one depicted in Fig. \ref{fig10}c, add to the oscillation presented 
in the bottom panel of Fig. \ref{fig9}. 

\begin{figure}[t]
\includegraphics[width=\columnwidth,clip=true]{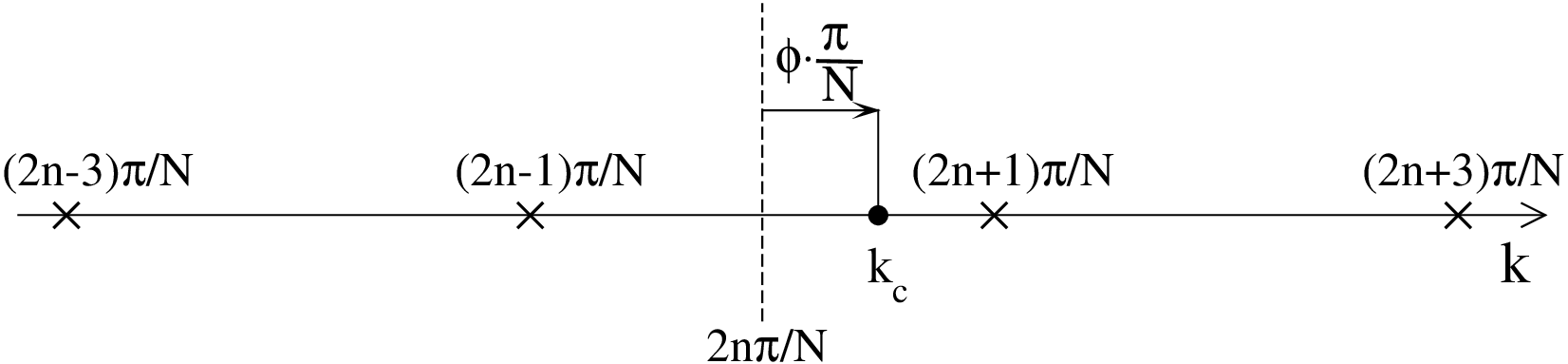}
\caption{Schematic plot illustrating  the shift 
$\phi$ from (\ref{An_phi}). Crosses show discretized momenta, 
the dot denotes position of $k_c$ (\ref{An_kc}), the dashed line goes through the
mid point between two discretized momenta being closest to $k_c=(2m+\phi)\pi/N$.
Note that $\phi>0$ ($\phi<0$)  when $k_c$ is to the right (left) 
of the dashed line. Naturally,  when $\phi$ approaches $\pm1$, $k_c$ collapses onto one of 
the discretized $k$'s. 
}
\label{fig11}
\end{figure}

\noindent Quantitatively,  
we introduce the parameter $\phi$  measuring distance of $k_c$ from the mid point between two 
momenta $k$ (\ref{Fprod_k}):
\begin{equation}
\phi=  \frac{N k_c}{\pi}~{\rm modulo}~ 2\in (-1,1].
\label{An_phi}
\end{equation}
Its definition is illustrated in detail in Fig. \ref{fig11}.
The examples of $\phi=1$, $0$ and $\frac{1}{2}$  are presented in Fig. \ref{fig10}. 
To extract the oscillatory part we focus on contribution to fidelity 
coming from momenta around  $k_c$. 
Taking the lowest order term of the Taylor expansion of (\ref{Fprod_fk})
around $k_c$ we obtain 
\begin{equation}
f_k \approx \alpha|k-k_c|
\label{swinka}
\end{equation}
where $\alpha=1/|\delta|(1-c^2)$ is introduced for convenience [compare with (\ref{fk_expansion})].
 Without going
into details, we define $k_{\rm cutoff}$ estimating the largest
$|k-k_c|$ for which (\ref{swinka}) provides a reasonable approximation to 
the exact expression for $f_k$.
Following
notation introduced in Fig. \ref{fig11}, we easily find that  
$f_k$ coming from the $m$-th momenta to the left (plus sign) or right (minus sign)  of $k_c$
 equals 
$$
f_k \approx \alpha\frac{\pi}{N}(2m-1\pm\phi).
$$
\noindent Now we can find how the asymmetry of  momenta $k$ around $k_c$ 
modifies fidelity. We do it by factoring out the contributions coming from 
the $\phi=0$ case where no asymmetry is present.
Namely, we calculate 
\begin{equation}
\frac{ \prod_k f_k }{ \prod_k f_k|_{\phi=0} } \simeq 
\prod_{m=1,2,\ldots}^{\frac{N}{\pi}k_{\rm cutoff}}
\left[ 1-\frac{\phi^2}{(2m-1)^2}\right] \overset{N\rightarrow \infty}{=} 
 \cos\frac{\pi\phi}{2}. 
 \label{kaszel}
\end{equation}
The last equality is reached  when 
$N k_{\rm cutoff}/\pi\gg1$. The change of the product into 
the cosine  can be found in  \cite{Ryzik}. 

To use  (\ref{kaszel}), we notice that for $\phi=0$ fidelity 
is given by $2{\cal F}_{\rm smooth}$ (\ref{fid_isingXX}). Thus, for
any $\phi$ it is given by $2\cos(\pi\phi/2){\cal F}_{\rm smooth}$ as long as $|c|<1$.
The origin of the prefactor of $2$ which multiplies ${\cal F} _{\rm smooth}$ is the same as in Sec. \ref{Sec_Ising}.
We have to correct $\ln{\cal F}_{\rm smooth}$ by a subleading term of $\ln 2$ to account 
for logarithmic singularities of $\ln f_k$ near $k_c$ (\ref{log_shift}). 
Note that for $\phi=0$ the displacement of momenta $k$ around $k_c$ is equivalent
to displacement of momenta $k$ around $k_c=0$ for the Ising model -- Sec. \ref{Sec_Ising}.
Thus, we can write our final prediction for fidelity in the following 
form 
\small
\begin{equation}
{\cal F} = \left \{
\begin{array}{lr} 
 2 \left|\cos\left(\frac{k_c N}{2}\right) \right| 
\exp\left[-N |\delta| 2 A(c) + N{\cal O}(\delta^2)\right]  & {\rm for}~ |c| < 1,  \\
\exp\left[-N |\delta| 2 A(c) +  N{\cal O}(\delta^2)\right] & {\rm for}~|c|\ge1,
\end{array}
\right. 
\label{FAnfinal}
\end{equation}
\normalsize
where  (\ref{An_phi})  was employed to simplify the oscillating factor.
The factor of two and sinusoidal shape of oscillations is reproduced by numerics when the system size 
$N$ is much larger  than the larger of the two correlation lengths
proportional to $1/|\delta (1-|c|)|$ (compare to Sec. \ref{Sec_Ising}).

\begin{figure}
\includegraphics[width=\columnwidth,clip=true]{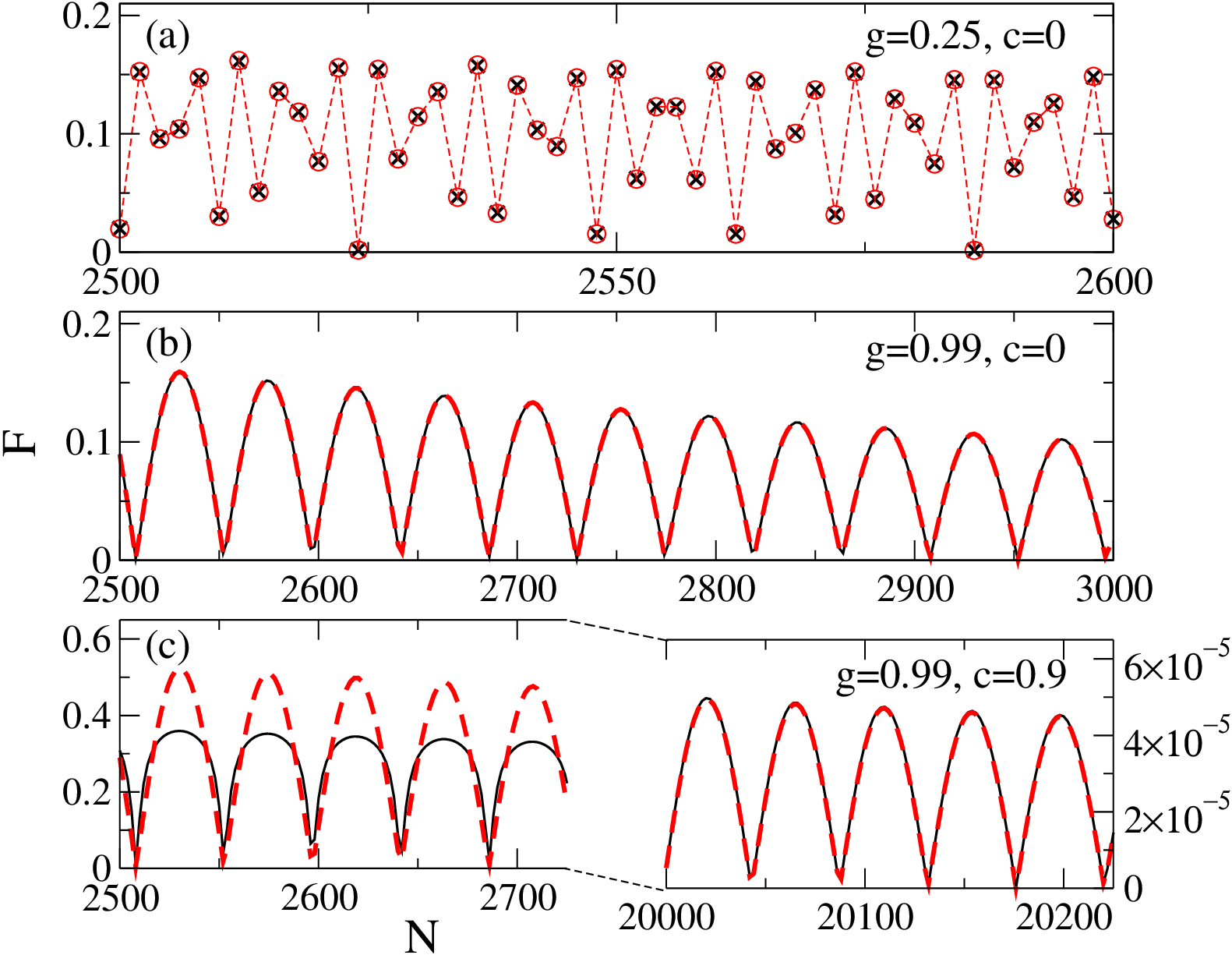}
\caption{(Color online). 
Plot of fidelity for the transition across the $\gamma=0$ line for $\delta=0.002$ and various parameters $g$ and $|c|<1$. 
Black line and crosses show numerical results. Red dashed line and circles 
show  theoretical prediction (\ref{FAnfinal}).}
\label{fig12}
\end{figure}

The oscillating part of  (\ref{FAnfinal}) is illustrated in Fig. \ref{fig12}. 
As we see in panels (a) and  (b), an agreement between
(\ref{FAnfinal}) and numerics is remarkable. The limits of applicability of 
(\ref{FAnfinal}) are illustrated in panel (c) of Fig. \ref{fig12}. The left 
part of this panel shows discrepancies resulting from application of
(\ref{FAnfinal}) to systems whose size $N$ is not much larger than the 
larger correlation length  ($\sim 1/|\delta (1-|c|)|$) -- still notice that the period of oscillations is predicted correctly. The right panel
reveals that for large enough  system sizes, and the very same parameters, the
perfect agreement is recovered.
We note in passing that those oscillations will survive in the 
``small system'' limit making the fidelity susceptibility approach
  (\ref{fid_sus}), which assumes that fidelity stays close to unity, inapplicable to this problem.

We expect that the same mechanism as described above should be able to explain oscillations of fidelity 
(peaks in fidelity susceptibility) observed  in the gapless phase of  the  Kitaev model \cite{FS_Kitaev1}.

Finally, it turns out that we can extend the results of this section. 
The condition  $|\delta|, |\epsilon| \ll \sqrt{1-g^2}$, which we have been using until now and which is necessary to justify approximations used above, turns out to be too strong.
Numerics shows that equation (\ref{FAnfinal}) holds for any $|\delta|, |\epsilon| \ll 1$ and $-1 < g < 1$ when $N$
is large enough.

\subsection{On the g=1 critical line}
\label{Sec_NON}
In this paragraph we follow the path C from Fig. \ref{fig1}, i.e., 
we substitute into (\ref{FXY}) 
\begin{equation}
\gamma_{1,2}=\epsilon\pm\delta, \ \ g_{1,2}=1.
\label{Cpath}
\end{equation}
This choice of the path 
is much different from what  we have considered until now. Indeed, both ground
states are calculated at points on the XY phase diagram that lie on the critical line.
In particular, it means that the 
correlation length at any of the points is of the order of the system size
$N$.

We start by discussing fidelity  calculated away from the multicritical point 
($g=1,\gamma=0$). This requires  
$$|\delta| \ll |\epsilon| \le 1.$$ 
Notice that $\epsilon$  is {\it not} assumed to be small here.

\noindent The exact expression for fidelity is given by substituting 
(\ref{Cpath}) into (\ref{Fprod}-\ref{Fprod_qk}). It can be simplified as follows.
Since 
\begin{eqnarray}
\label{98}
p_k&=&(1-\cos k)^2+ (\epsilon^2-\delta^2) \sin^2 k,\\
q_k&=&2\delta (1-\cos k ) \sin k, \label{89}
\end{eqnarray}
for the C path,  we see that $p_k$ is positive and 
$q^2_k/p^2_k\le \delta^2/(\epsilon^2-\delta^2)$ is small. 
Expanding $\ln f_k$  in $q_k/p_k$ we obtain 
$$
\ln f_k \simeq - \frac{q_k^2}{8 p_k ^2}.
$$
It can be analytically integrated over $k$ from $0$ to
$\pi$  (\ref{Fintegral}), and then expanded in $\delta$ to yield 
\begin{equation}
\frac{\ln {\cal F}}{N} \simeq - \frac{\delta^2}{8 |\epsilon| (1+|\epsilon|)^2} + {\cal O}(\delta^4).
\label{fid_critical}
\end{equation}

\begin{figure}[t]
\includegraphics[width=\columnwidth,clip=true]{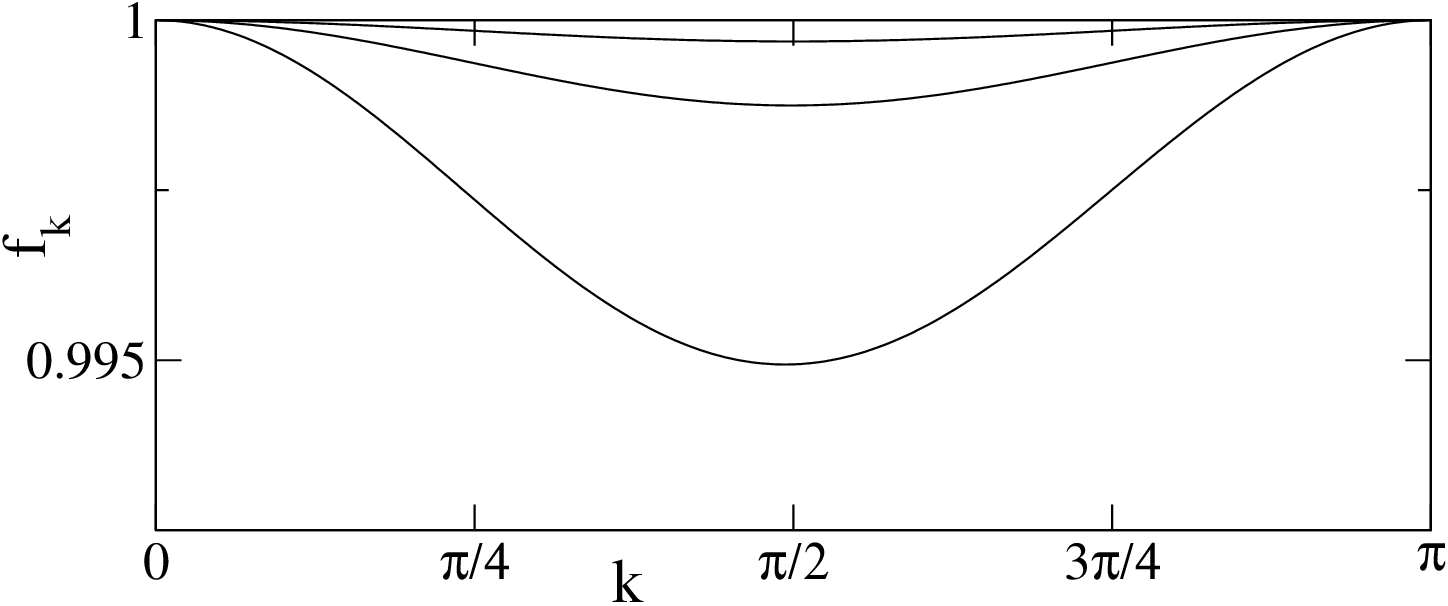}
\caption{$f_k$ along the Ising critical line (\ref{Cpath}) for $\epsilon=1$ and 
$\delta=0.05,0.1,0.2$ (top to bottom). The minimum of $f_k$
is reached for $k=\arccos\left( \frac{1-\epsilon^2+\delta^2}{1+\epsilon^2-\delta^2}\right)\simeq\arccos\left( \frac{1-\epsilon^2}{1+\epsilon^2}\right)$.
}
\label{fig13}
\end{figure}
\noindent 
Equation  (\ref{fid_critical}) provides an interesting result showing that the universal part of
fidelity, present in Secs. \ref{Sec_Ising} and \ref{Sec_XX}, is absent here.
Therefore, only the nonuniversal part proportional to quadratic displacement 
of the ground states is present.

\noindent Looking at fidelity from the momentum angle, Fig. \ref{fig13}, we notice that $f_k$'s are very
different here from what we have studied in the previous sections.
Not only $f_k$'s stay close to unity for all momenta, but also 
the main contribution to fidelity comes in general from large $k$'s that are not associated
with disappearance of the gap: the gap (\ref{energy_gap}) closes at
$k_c=0$ on this line. 
This shows that the ground states used to calculate fidelity  differ only by the nonuniversal contribution 
away from the multicritical point.
This is to be expected from the renormalization group perspective because 
we calculate fidelity between states which differ only  by an {\it irrelevant}
variable.

Now we would like to show what happens near the multicritical point and so we
assume that 
$$
|\epsilon|,|\delta| \ll 1,
$$
and use the familiar parameterization $\epsilon = c |\delta|$. The dominant contribution to fidelity comes 
now from momenta around $k_c=0$.

\noindent
Approximating 
$p_k\approx k^2(k^2/4+\epsilon^2-\delta^2)$ and $q_k\approx k^3\delta$ from
(\ref{98},\ref{89}) and using the same methods  as in Sec. \ref{Sec_Ising}, we obtain  
\begin{equation}
\ln {\cal F} = \left \{
\begin{array}{cc} 
\begin{split}
& -N \left[   |\delta| 2 A(c) + {\cal O}(\delta^2)\right] + \ln 2 /2  & {\rm for}~ |c|<1,  \\
&  -N \left[  |\delta| 2 A(c) +  {\cal O}(\delta^2)\right] & {\rm for}~|c|\ge1,
\end{split}
\end{array}
\right. 
\label{FAnfinalMCP}
\end{equation}
where the scaling function $A(c)$ is given by (\ref{Ac}). 

\noindent This is a pretty interesting result showing that despite the fact that 
the ground states entering fidelity are calculated on the critical line,
there is a universal contribution to fidelity. This contribution 
can be traced back to the multicritical point: the dominant critical point
on the $g=1$ line. Evidence for that is twofold. First, the universal 
result (\ref{FAnfinalMCP}) appears as we approach the multicritical point.
Second, the scaling exponent $\nu$, that can be extracted from
(\ref{FAnfinalMCP}) by comparing it to (\ref{near_F}), equals $1$.
It matches the predicted value for this multicritical point along the 
$g=1$ line/direction \cite{Mukherjee2011}. 
It should be also mentioned that the concept of  the dominant critical point 
has been introduced in \cite{Deng2008} in the context of quench dynamics.

Finally, it is interesting to see how (\ref{FAnfinalMCP}) crosses over into 
(\ref{fid_critical}) as we go  away from the multicritical point (the limit of $|c|\gg1$).
Assuming that $|\delta|\ll |\epsilon| \ll 1$, equation (\ref{FAnfinalMCP}) can be simplified to
$$
\frac{\ln {\cal F}}{N} \simeq - \frac{\delta^2}{8 |\epsilon|},
$$
which matches (\ref{fid_critical}) in the leading order in $\epsilon$. Note,
however, that one cannot apply (\ref{FAnfinalMCP}) to arbitrary distances 
away from the critical point because it is valid for $|\epsilon|=|c\delta|\ll1$ only. 
For larger $\epsilon$'s non-universal contributions 
dominate fidelity (\ref{fid_critical}).

\subsection{Near $\gamma$=0 and g=1: multicritical point}
\label{Sec_Lif}
In this paragraph we follow the path D from Fig. \ref{fig1}, i.e., 
we substitute into (\ref{FXY}) 
\begin{equation}
g_{1,2}=1+\epsilon\pm \delta, \ \gamma_{1,2}=\alpha(\epsilon\pm \delta), \ \epsilon=c|\delta|,
\label{Dpath}
\end{equation}
where $0<\alpha<\infty$ is the slope of our path. 
We restrict our studies to $g_{1,2}\ge1$ and $\gamma_{1,2}\ge0$: both ground states used to calculate 
fidelity are obtained on the paramagnetic side (Fig. \ref{fig1}). This requires
$\epsilon \ge |\delta|$ or equivalently $c\ge1$. Fidelity
calculated along this path is ``influenced'' by the multicritical point at
$g_c=1$ and $\gamma_c=0$.

\begin{figure}[t]
\includegraphics[width=\columnwidth,clip=true]{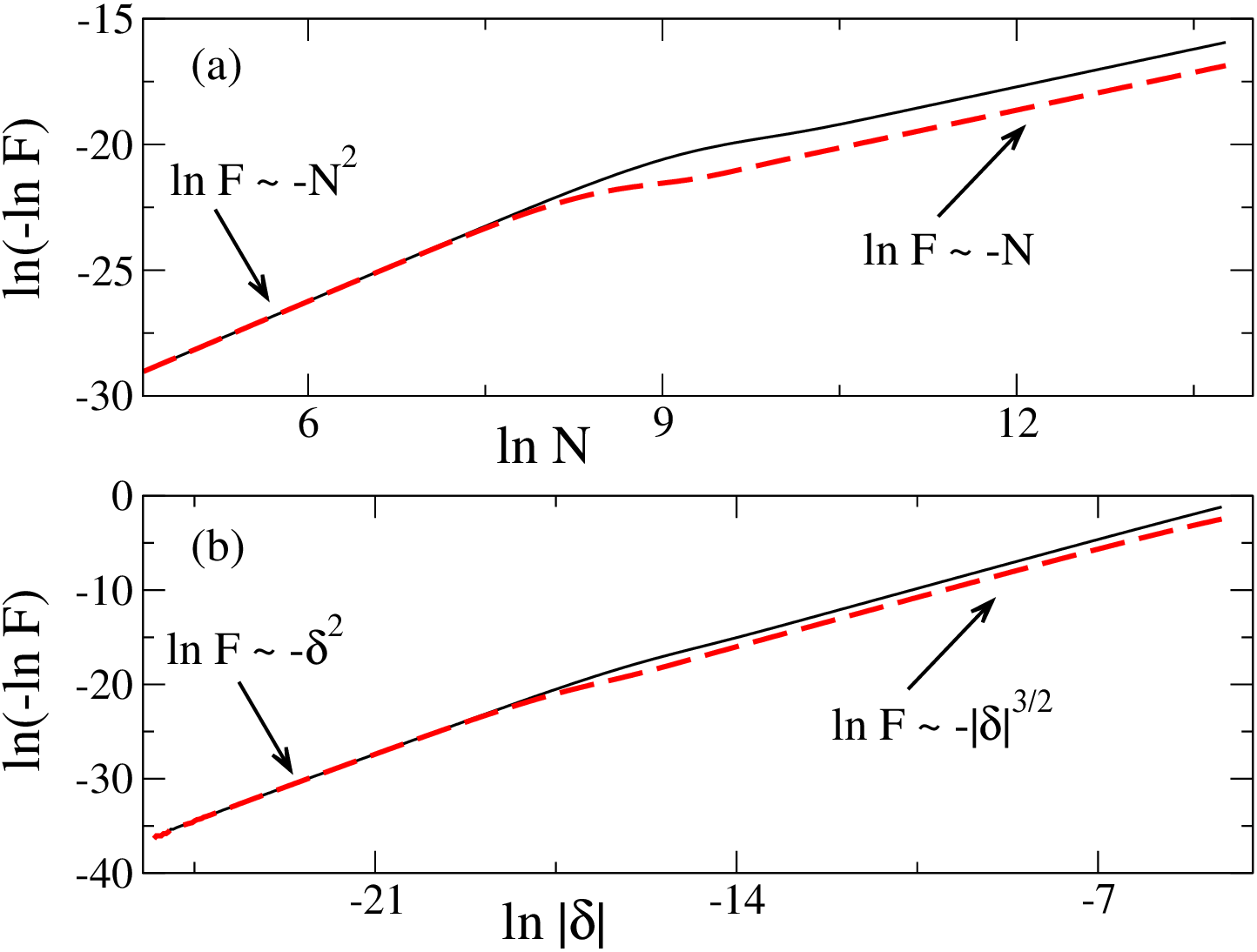}
\caption{
(Color online). 
Transition from the ``small system'' limit to the thermodynamic limit near the multicritical
 point located at  $g=1$ and $\gamma=0$. (a): the transition as a function of 
the system size $N$ at fixed $\delta=10^{-8}$.
 (b): the transition as a function of parameter difference $\delta$ at fixed $N=3000$.
In both plots solid (black) line shows fidelity for $c=1$, i.e., 
${\cal F}=|\langle 1,0|1+2\delta,2\delta\rangle|$, while 
the dashed (red) line shows the $c=5$ case, i.e., 
${\cal F}=|\langle 1+4\delta,4\delta|1+6\delta,6\delta\rangle|$ (\ref{Dpath}).  Note that
$\alpha=1$ is assumed here.
}
\label{fig14}
\end{figure}

The condition for the crossover between the ``small system'' limit  and the thermodynamic limit
near the multicritical point reads (\ref{tlimit})
\begin{equation}
N|\delta|^\nu\sim1.
\label{Lif_T}
\end{equation}
As we
deal here with a multicritical point that is characterized by more than one
divergent  length scale, it is important to carefully verify 
the above prediction to make sure that the relevant $\nu$
is used in  (\ref{Lif_T}).

The crossover is  illustrated in Fig. \ref{fig14}, where 
distinct scalings of fidelity with either the system size $N$ or the
parameter difference $\delta$ are easily observed.
In Fig. \ref{fig14}a
the parameter difference $\delta$ is kept constant and the system size $N$ is varied.
We see that for small system sizes $\ln{\cal F} \sim -N^2$, while for large
ones $\ln{\cal F}\sim -N$. In Fig. \ref{fig14}b the system size $N$
is kept constant, while the parameter shift $\delta$ is varied. Again two regimes
appear: for small $\delta$ we have $\ln{\cal F}\sim-\delta^2$, while
for larger $\delta$ we get $\ln{\cal F}\sim-|\delta|^{3/2}$.

\begin{figure}
\includegraphics[width=\columnwidth,clip=true]{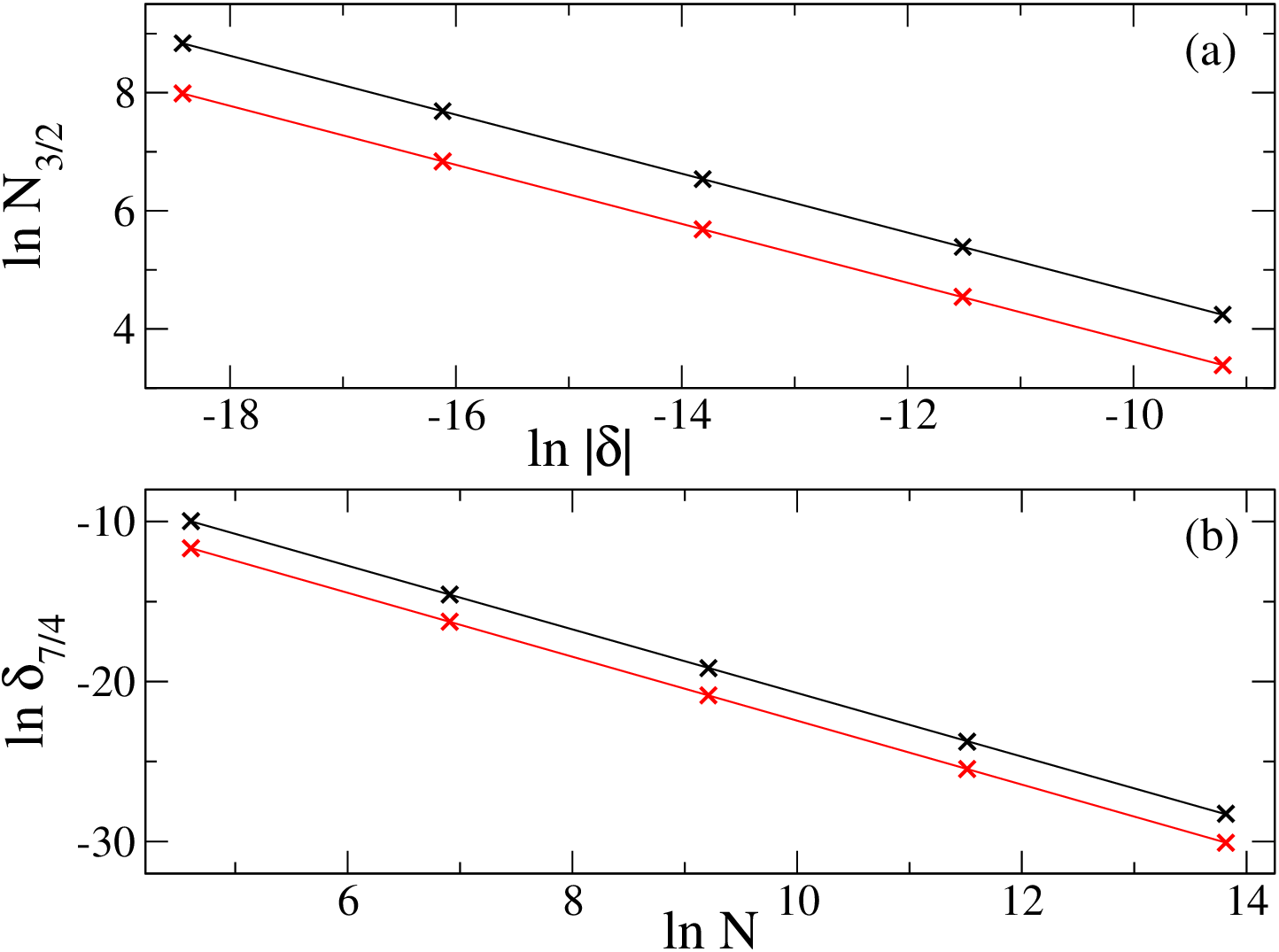}
\caption{(Color online).   Numerical study of the crossover condition 
(\ref{Lif_T}). Crosses come from numerics (see text for details), while 
straight lines are linear fits \cite{Fit}. In both panels upper (lower) data sets
correspond to $c=1$ ($c=5$). 
Panel (a): we fit $a$ and $b$
coefficients of $\ln N_{3/2}= a + b \ln \delta$. 
For $c=1$ ($c=5$) we obtain 
 $b=-0.499\pm 0.0003$  
 ($b=-0.4992\pm 0.0005$). 
Panel (b): we fit $a$ and $b$
coefficients of $\ln \delta_{7/4}= a + b \ln N$. 
For $c=1$ ($c=5$) we obtain  
$b=-1.989\pm 0.005$  
($b=-1.9987\pm 0.0008$). 
 The numerics for these plots is
 done with $\alpha=1$. The fitting coefficient $a$ is not listed as it is 
 of minor interest.
}
\label{fig15}
\end{figure}

The location of the crossover can be studied in exactly the same 
way as in Sec. \ref{Sec_Ising}. Briefly, 
as the system size $N$ is increased in Fig. \ref{fig14}a,  
the slope of $\ln(-\ln{\cal F})$  
changes smoothly from $2$ (corresponding to $\ln{\cal F}\sim -N^{2}$) to $1$
(corresponding to $\ln{\cal F}\sim -N$). The crossover region
between the two limits is centered around $N=N_{3/2}$ where the local slope equals $3/2$.
By repeating the calculation from Fig. \ref{fig14} for various $\delta$'s 
we have numerically obtained $N_{3/2}(\delta)$.
A power-law fit described in Fig. \ref{fig15}a reveals 
that $N_{3/2}\sim 1/\sqrt{|\delta|}$ supporting $\nu=1/2$ in (\ref{Lif_T}). Similar analysis can be performed 
on data from Fig. \ref{fig14}b. Indeed, now we look at 
$\delta_{7/4}(N)$ such that the slope of $\ln(-\ln{\cal F})$ 
equals $7/4$, i.e., is half-way between $2$ and $3/2$ observed on both 
``ends'' of Fig. \ref{fig14}b. Fig. \ref{fig15}b, where the power-law
fit is performed, shows that  $\delta_{7/4} \sim N^{-2}$ again pointing to
$\nu=1/2$. Therefore,
we conclude that the crossover condition (\ref{Lif_T}) holds near the
multicritical point with $\nu=1/2$ on the paramagnetic side.

\begin{figure}
\includegraphics[width=\columnwidth,clip=true]{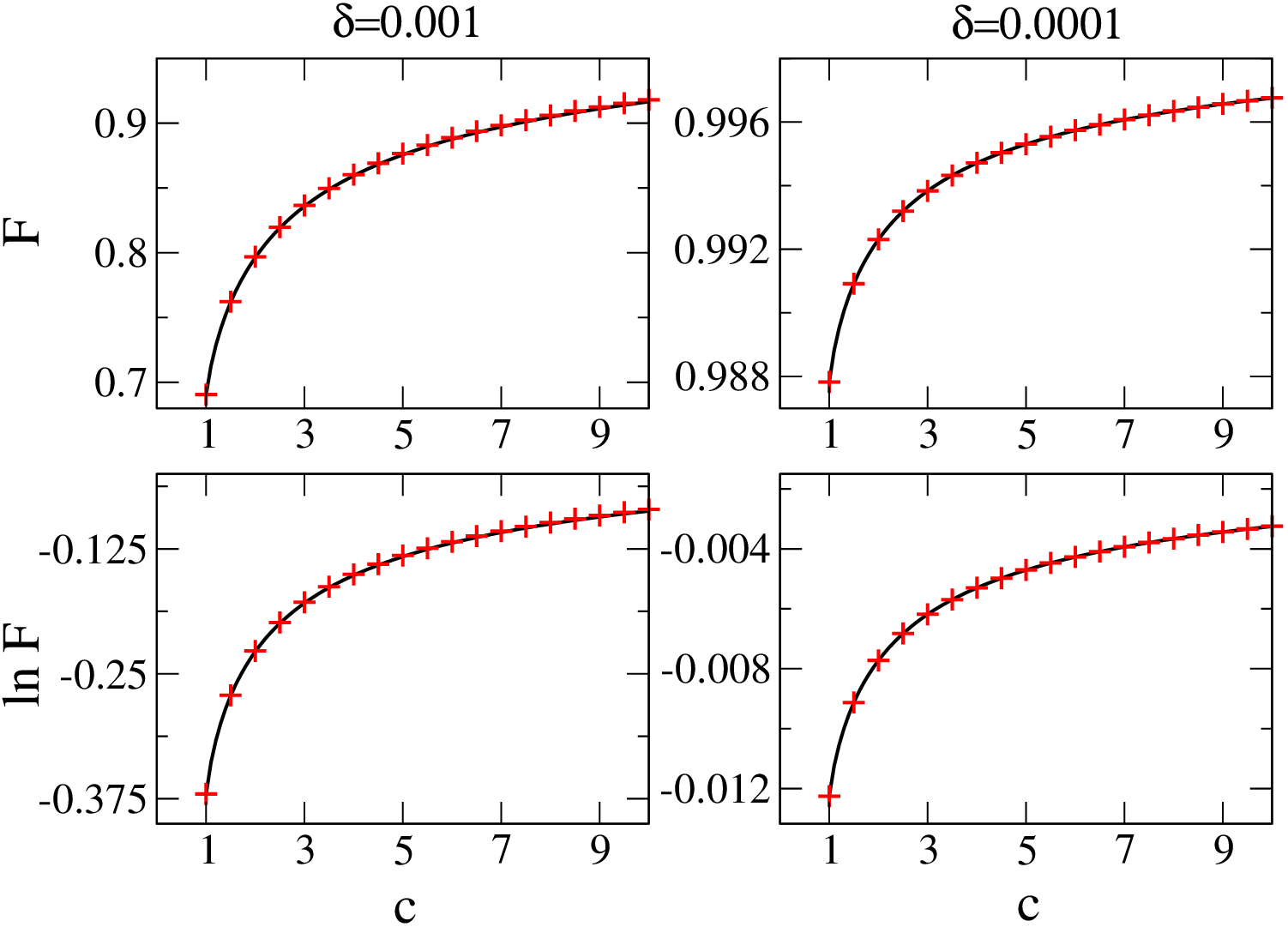}
\caption{(Color online). 
Comparison between numerics (solid lines) and analytics [pluses given by (\ref{Lif_fidel})]. 
In all panels $N=10^5$ and $\alpha=1$. The parameter shift $\delta$ is chosen
to be small enough to keep the ${\cal O}(|\delta|^{5/2})$ corrections to (\ref{Lif_fidel}) negligible.
}
\label{fig16}
\end{figure}

We focus on the thermodynamic limit again. 
Substituting (\ref{Dpath}) into (\ref{Fprod_pk},\ref{Fprod_qk}) we obtain 
\begin{eqnarray*}
p_k&=&(1+\epsilon-\cos k)^2-\delta^2 +\alpha^2 (\epsilon^2-\delta^2) \sin^2 k,  \\
q_k&=&2 \alpha \delta (1-\cos k ) \sin k. 
\end{eqnarray*}
When both points -- $(g_1,\gamma_1)$ and $(g_2,\gamma_2)$  (\ref{Dpath}) -- are on the paramagnetic side,  then $p_k >0$ and
$q^2_k/p^2_k < \alpha|\delta|/2 \ll 1$ 
is a small parameter in which we expand $\ln f_k$.
This brings us again  to
\begin{equation}
\ln f_k \simeq -\frac{q_k^2}{8 p_k^2}.
\label{flu}
\end{equation}
Integrating (\ref{flu}) over $k$ from $0$ to $\pi$ (\ref{Fintegral}),
and then expanding the resulting expression in $\delta$ we get  
\begin{equation}
\frac{\ln{\cal F}}{N} = - |\delta|^{3/2} \alpha^2 A_{\rm MCP}(c) +  \delta^2 \alpha^2 /4 + {\cal O}(|\delta|^{5/2}).
\label{Lif_fidel}
\end{equation}
The scaling function for the multicritical point is given by
$$
A_{\rm MCP}(c)=\frac{(c+1)^{3/2}(3-2c)+(c-1)^{3/2}(3+2c)}{16 \sqrt{2}},
$$
which is valid for $c\ge 1$.
This result is illustrated in Fig. \ref{fig16}, where small enough  $\delta$ has been chosen to 
keep the terms ${\cal O}(|\delta|^{5/2})$ negligible. 
We  assume such choice of 
$\delta$ from now on.  The result (\ref{Lif_fidel}) is  interesting for several reasons.

First, the exponent $3/2$ does not fit the expected $d\nu$ value 
from (\ref{near_F}): neither $\nu=1/2$ nor $\nu=1$, both considered
near the multicritical point \cite{Mukherjee2011}, agree here.
Explanation of that anomaly is beyond the scope of this work.

Second, due to remarkable simplicity of (\ref{Lif_fidel}) we can analyze in detail
the interplay between the nonanalytic and the subleading contribution to  fidelity.
The nonanalytic term dominates over the subleading one when 
$$
|\delta|^{3/2} \alpha^2 A_{\rm MCP}(c) \gg 
\delta^2 \alpha^2 /4 \Rightarrow |\delta|\ll 16 A_{\rm MCP}(c)^2\le1/4.
$$
Additionally, the ``thermodynamic limit'' condition $N\sqrt{\delta}\gg1$ must be satisfied so 
that (\ref{Lif_fidel}) holds. If these two conditions
are fullfield, $\ln{\cal F}\approx -N |\delta|^{3/2} \alpha^2 A_{\rm MCP}(c)$. This does not, however, 
guarantee that ${\cal F}$ is well approximated by $\exp(-N |\delta|^{3/2} \alpha^2 A_{\rm MCP}(c))$. 
Indeed,
the latter requirement puts an additional bound on the parameter shift $\delta$: 
$$
N \delta^2 \alpha^2 /4 \ll 1 \Rightarrow \exp(N \delta^2 \alpha^2 /4) \approx 1.
$$ 
All these conditions can be simultaneously satisfied.

Third, 
equation (\ref{Lif_fidel}) shows that fidelity near the multicritical point -- at least in the paramagnetic phase discussed here -- does not have to be  
small in the thermodynamic limit. This is in stark contrast to what we found  near the Ising critical point, where 
the thermodynamic-limit condition (\ref{Ising_thermo}) implied smallness of 
${\cal F}$ (\ref{z}). Here, we can have $N\sqrt{|\delta|}\gg1$ and still 
$\exp(-N |\delta|^{3/2} \alpha^2 A_{\rm MCP}(c) + N \delta^2 \alpha^2 /4)$
 anywhere between $0$ and $1$.
 
\begin{figure}
\includegraphics[width=\columnwidth,clip=true]{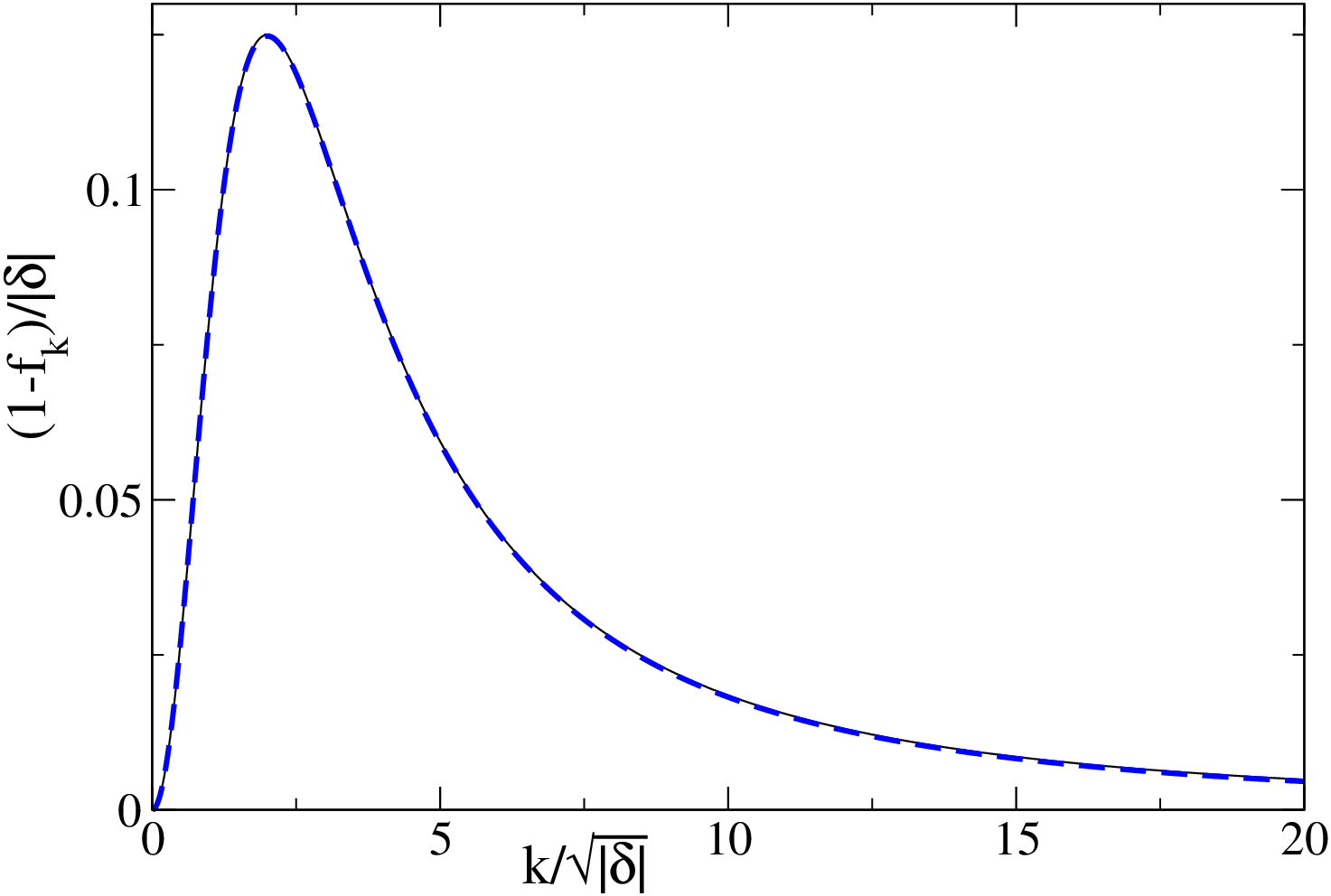}
\caption{(Color online). We show here from the momentum space angle that
$\ln{\cal F}\sim -|\delta|^{3/2}$ (\ref{Lif_fidel}). We plot the rescaled $f_k$ functions for 
$c=1$ and $\alpha=1$:
one for $\delta=0.001$ (solid black line) and the other for $\delta=0.01$
(dashed blue line). As the two curves overlap very well,  we 
see that  $(1-f_k)/|\delta| = h(k/\sqrt{|\delta|})$, where $h$
is some non-singular function. 
This  implies that $\ln {\cal F} \sim \int dk \ln f_k  \sim -|\delta|^{3/2}$ ($f_k \approx 1$ so we approximate 
$\ln f_k \simeq f_k-1$). This confirms leading scaling of (\ref{Lif_fidel})
with $\delta$ and independently confirms the condition (\ref{Lif_T}) with
$\nu=1/2$. 
We also note that these rescalings work in the limit of $\delta\to0$. 
}
\label{fig17}
\end{figure}

Fourth, 
 similarly as in the Ising  chain one can easily derive from (\ref{Lif_fidel}) transition to the analytic limit in $\delta$ 
 far away from the critical point. Considering $\epsilon\gg|\delta|$ or simply $c\gg 1$ 
 (but still $\epsilon = c|\delta|\ll1$), 
  $A_{\rm MCP}(c) \simeq 5/32\sqrt{2c}$ and so 
\begin{equation}
{\cal  F} \simeq \exp\left(-\delta^2 N \frac{5\alpha^2} {32 \sqrt{2\epsilon}} \right).
\label{XYaway}
\end{equation}
This is again a new result. If we expand (\ref{XYaway}) assuming that the argument under the exponent is small, 
we get a fidelity susceptibility-like expression in the thermodynamic limit
$$
{\cal F}\simeq 1-\delta^2\chi_F/2, \ \ \chi_F \approx N\frac{5\alpha^2}{16\sqrt{2\epsilon}}.
$$
The above equation can be also obtained 
when we start with (\ref{fid_sus}) and expand the resulting expression  
for $\chi_F$ in $\epsilon$. 
It predicts that  fidelity susceptibility away from the multicritical point in the paramagnetic phase 
scales as $\epsilon^{-1/2}$.

Finally, we mention that  one can find the $|\delta|^{3/2}$ scaling of $\ln{\cal F}$ 
by looking directly at $f_k$. This is explained in  Fig. \ref{fig17}.

\section{Extended Ising model}
\label{Sec_MPS}
In this section we consider the Hamiltonian \cite{Wolf2006}
\begin{eqnarray} 
\label{H_MPS}
\hat H&=&\sum_{n=1}^N \left[-2(1-g^2)\sigma_n^x \sigma_{n+1}^x -(1+g)^2 \sigma_n^z\right.
\nonumber\\ &&~~~~~~+\left.(1-g)^2 \sigma_n^x \sigma_{n+1}^z \sigma_{n+2}^x\right].
\end{eqnarray}
with periodic boundary conditions $\sigma_{n+N} = \sigma_n$.
Its ground state is given exactly by finite rank Matrix Product State \cite{Wolf2006}.
It has the critical point at $g_c =0$, which is characterized by the critical exponents $\nu=1$ and $z=2$. 
In contrast to the Ising model, neither 
entropy of entanglement nor the ground state energy is singular at this critical point.
Still, as we will see below, the basic features of that transition are captured by fidelity.
Introducing the notation $|g\rangle$ for the ground state of (\ref{H_MPS}), and
defining ${\cal F}$ as $|\langle g_1 | g_2\rangle|$ one obtains that 
\begin{equation}
{\cal F}= \frac{\left|(1+\sqrt{g_1 g_2})^N + (1-\sqrt{g_1 g_2})^N\right|}{\sqrt{(1+g_1)^N + (1-g_1)^N} \sqrt{(1+g_2)^N + (1-g_2)^N}}.
\label{fidel_MPS1}
\end{equation}
This is an exact expression from  \cite{Cozzini2007,Zhou2008}.
We are going to simplify this result to discuss it from the perspective relevant for our approach.

We follow notation from Sec. \ref{Sec_XYmodel} writing 
\begin{equation}
g_1 = \epsilon + \delta, \ g_2 = \epsilon - \delta, \ \epsilon=c|\delta|.
\label{dupa4}
\end{equation}
We stay close to the critical point so that $|\delta|,|\epsilon| \ll 1$ and 
 approximate (\ref{fidel_MPS1}) with hyperbolic functions
\begin{equation}
\label{fidel_MPS2}
{\cal F}\simeq \frac{\left| \cosh\left(N \sqrt{\epsilon^2-\delta^2 }\right)\right|}{\sqrt{\cosh\left(N  (\delta+\epsilon )\right) \cosh\left(N ( \epsilon -\delta )\right)}}. 
\end{equation}
This simple expression allows us to easily study the ``small system'' limit
and the thermodynamic limit.

First, we focus on  the ``small system'' limit, say $\delta\to0$ at fixed $N$ for simplicity.  
Expansion of (\ref{fidel_MPS2}) in $\delta$, the fidelity susceptibility approach (\ref{fid_sus}),
results in
\begin{equation}
{\cal F}\simeq 1- \frac{\delta ^2 N^2}{2} \left[\frac{1}{\cosh^2(\epsilon N)} +\frac{\tanh(\epsilon N)}{\epsilon N}\right].
\label{MPS_sus}
\end{equation}
This expression depends on two parameters: $N|\delta|$ and $N|\epsilon|$. 
As we keep the former small,  we can focus on its $N|\epsilon|$ dependence.
When $|\epsilon| N \ll 1$ then (\ref{MPS_sus}) simplifies to
$$
{\cal F}\simeq 1- \delta ^2 N^2
$$
reproducing expected scaling of fidelity susceptibility at the critical point 
\cite{Schwandt2009,ABQ2010,BarankovArXiv2009,deGrandi2010a,deGrandi2010b,PolkovnikovArXiv2010}.
In the opposite limit of $|\epsilon| N \gg 1$,  (\ref{MPS_sus}) can be written as  
\begin{equation}
{\cal F}\simeq 1- \delta ^2 N /2|\epsilon|.
\label{1234}
\end{equation}
This again agrees with the scaling predictions 
\cite{Schwandt2009,ABQ2010,BarankovArXiv2009,deGrandi2010a,deGrandi2010b,PolkovnikovArXiv2010}.
Finally, note that when (\ref{MPS_sus}) holds,  fidelity stays very close to unity.

Second, we concentrate on the thermodynamic limit, say $N\rightarrow\infty$ at fixed $\delta$,
where we obtain from (\ref{fidel_MPS2})
\begin{equation}
{\cal F} \simeq \left\{
\begin{array}{cl} 
\begin{split}
& 2   \left| \cos\left( |\delta| N \sqrt{1-c^2}\right)\right| \exp\left(-|\delta| N \right)& {\rm for}~|c| < 1,  \\
& \sqrt{2} \exp\left(-|\delta| N \right)& {\rm for} ~ |c|=1, \\
& \exp\left(-|\delta| N \left( |c|-\sqrt{c^2-1} \right)\right)& {\rm for} ~  |c| > 1.
\end{split}
\end{array}
\right. 
\label{fidel_MPS3}
\end{equation}

\noindent To be more precise, 
we mention that (\ref{fidel_MPS3}) provides a good approximation to the exact
result when $\max\left[ N|\epsilon+\delta|,N|\epsilon-\delta| \right]\gg1$, which is
a thermodynamic limit condition based on (\ref{tlimit}).

\noindent Similarly as for the XY model,  we  introduce the scaling function $A_{MPS}(c)$ as 
\begin{equation}
A_{MPS}(c) = \left\{
\begin{array}{c} 
\begin{split}
& 1 & {\rm for} ~ |c|  \le  1,  \\
& |c|-\sqrt{c^2-1}& {\rm for} ~   |c|>1.
\end{split}
\end{array}
\right.
\label{Ac_MPS}
\end{equation}
$A_{MPS}(c)$ is nonanalytic when $|c|=1$, i.e., 
when one of the states is exactly at the critical point, signaling the 
pinch point singularity
\cite{Zhou2008,Zhou2008_Vidal,Zhou2008_Ising}. As expected, 
this singularity is rounded off in finite systems.

\noindent When $|c| \gg 1$, or equivalently  $|\delta|\ll|\epsilon|\ll1$, 
then the scaling function $A_{MPS}$ approaches zero as $1/2|c|$ leaving us
with 
$$
{\cal F}\simeq\exp\left(-\delta^2 N / 2 | \epsilon |\right).
$$
This coincides with the fidelity susceptibility result (\ref{1234}) when the argument of the exponent is small,
but provides a different result otherwise.

Now we will study the origin of oscillations and prefactors appearing in (\ref{fidel_MPS3}) 
illustrating how the analytical  techniques from  Sec. \ref{Sec_XYmodel} 
can be applied to other models as well.
Hamiltonian (\ref{H_MPS}) can be diagonalized using exactly the same formalism as in Sec. \ref{Sec_XYmodel}. 
Indeed, the Jordan-Wigner transformation translates
(\ref{H_MPS}) into a chain of noninteracting fermions which can be  solved
using Fourier and Bogoliubov transformations.
We assume even $N$ and follow notation from \cite{JacekPRL2005} 
during diagonalization of (\ref{H_MPS}). 
The ground state lays then in a subspace with even number of quasiparticles. 
In that subspace the Hamiltonian (\ref{H_MPS})  is diagonalized to the form 
$\hat H=\sum_{k>0 } \epsilon_k \left( \gamma_k^\dagger \gamma_k + \gamma_{-k}^\dagger \gamma_{-k}  -1 \right)$,
where
$\gamma_k$ are the fermionic annihilation operators  and the energy gap is given by
$$
\epsilon_k = 4 \left(1+g^2 - (1-g^2) \cos k  \right).
$$
It reaches zero at the critical point $g_c=0$ for $k_c=0$.
For small $k$ and $g$ we can approximate $\epsilon_k \approx 8 g^2 + 2 k^2$, 
which confirms the critical exponents $\nu=1$ and $z=2$ (at the critical point 
the gap scales as $k^z$, while near the critical point it closes as $|g-g_c|^{z\nu}$).

\begin{figure}
\includegraphics[width=\columnwidth,clip=true]{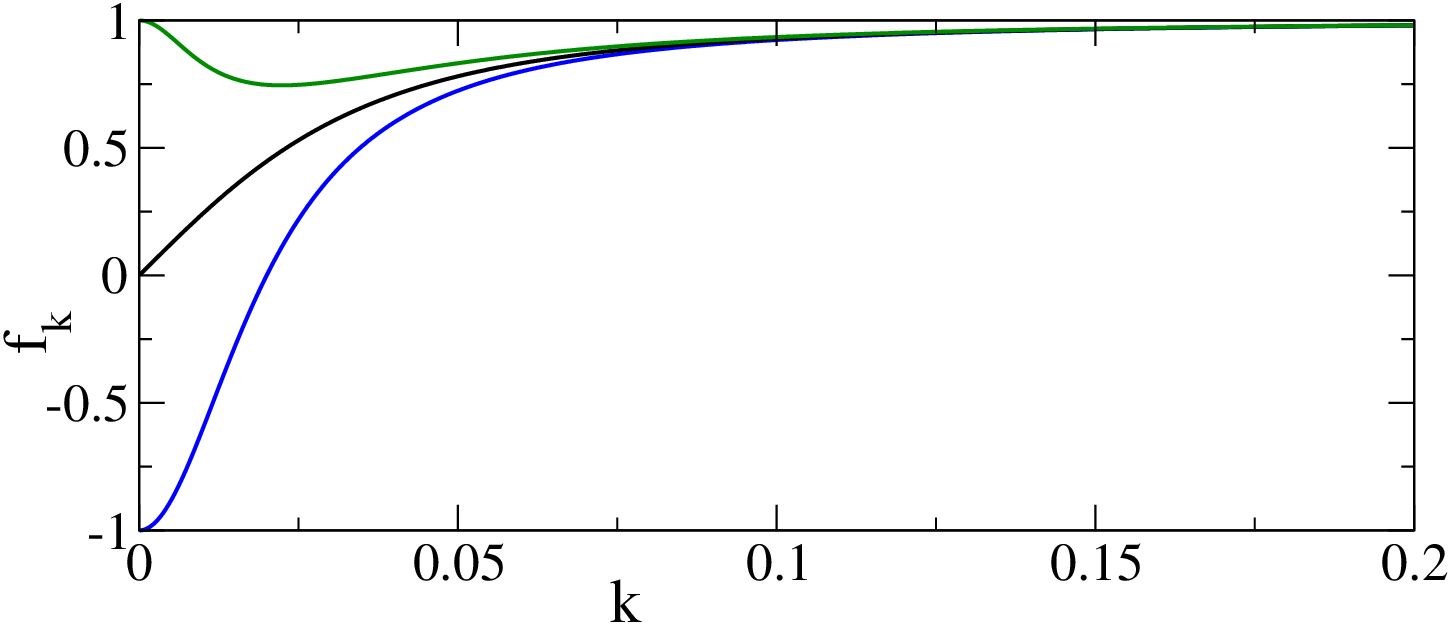}
\caption{(Color online). The lines correspond to $f_k$ plotted  for $\delta=0.01$ (\ref{Fprod_fk_MPS}).
From bottom to top  $c$ equals  $0$, $1$ and $1.5$,  respectively. For $c=0$
the ground states entering fidelity are calculated on the opposite sides of the critical
point, for $c=1$ one of them is obtained  at the critical point, while for 
$c=1.5$ they are both obtained on  the same side of the critical point.}
\label{fig18}
\end{figure}

Fidelity is given by 
${\cal F} = \prod_{k>0}\left|\cos(\theta^1_k/2-\theta^2_k/2)\right|$, where this time $\theta^i_k$'s are defined by
$$
\tan \theta^i_k = \frac{2 (1-g_i^2) \sin k - (1-g_i)^2 \sin 2k}{(1+g_i)^2-2 (1-g_i^2) \cos k + (1-g_i)^2 \cos 2k}.
$$
After some algebra we arrive at 
\begin{eqnarray}
&{\cal F}&=\prod_{k>0} |f_k|,  \label{Fprod_MPS} \\ 
&k& = (2n+1)\pi/N, \  n = 0,1,\ldots,N/2-1,\label{Fprod_k_MPS}\\ 
&f_k&= \frac{p_k}{\sqrt{p_k^2+q_k^2}},\label{Fprod_fk_MPS}\\
&p_k& =  1+g_1 g_2 -(1- g_1 g_2) \cos k, \label{Fprod_pk_MPS}\\
&q_k& =  (g_1 - g_2) \sin k. \label{Fprod_qk_MPS}
\end{eqnarray}
For the sake of presentation we allow $f_k$ to take negative values here and use the  parameterization (\ref{dupa4}).
The prototypical behavior of $f_k$  is shown in Fig. \ref{fig18}.  

We expand $p_k$ and $q_k$ around $k_c=0$ and for $|\epsilon|,|\delta|\ll1$ we get
$p_k\approx k^2/2 +  2(\epsilon^2 - \delta^2)$ and $q_k\approx 2 \delta k$.
Thus we can approximate
\begin{equation}
\ln|f_k| \approx \frac{1}{2} \ln \left( \frac{(k^2/4 +  \epsilon^2 - \delta^2)^2}{(k^2/4 +  \epsilon^2 - \delta^2)^2+(\delta k)^2}  \right).
\label{tildefk_MPS}
\end{equation}

\begin{figure}
\includegraphics[width=\columnwidth,clip=true]{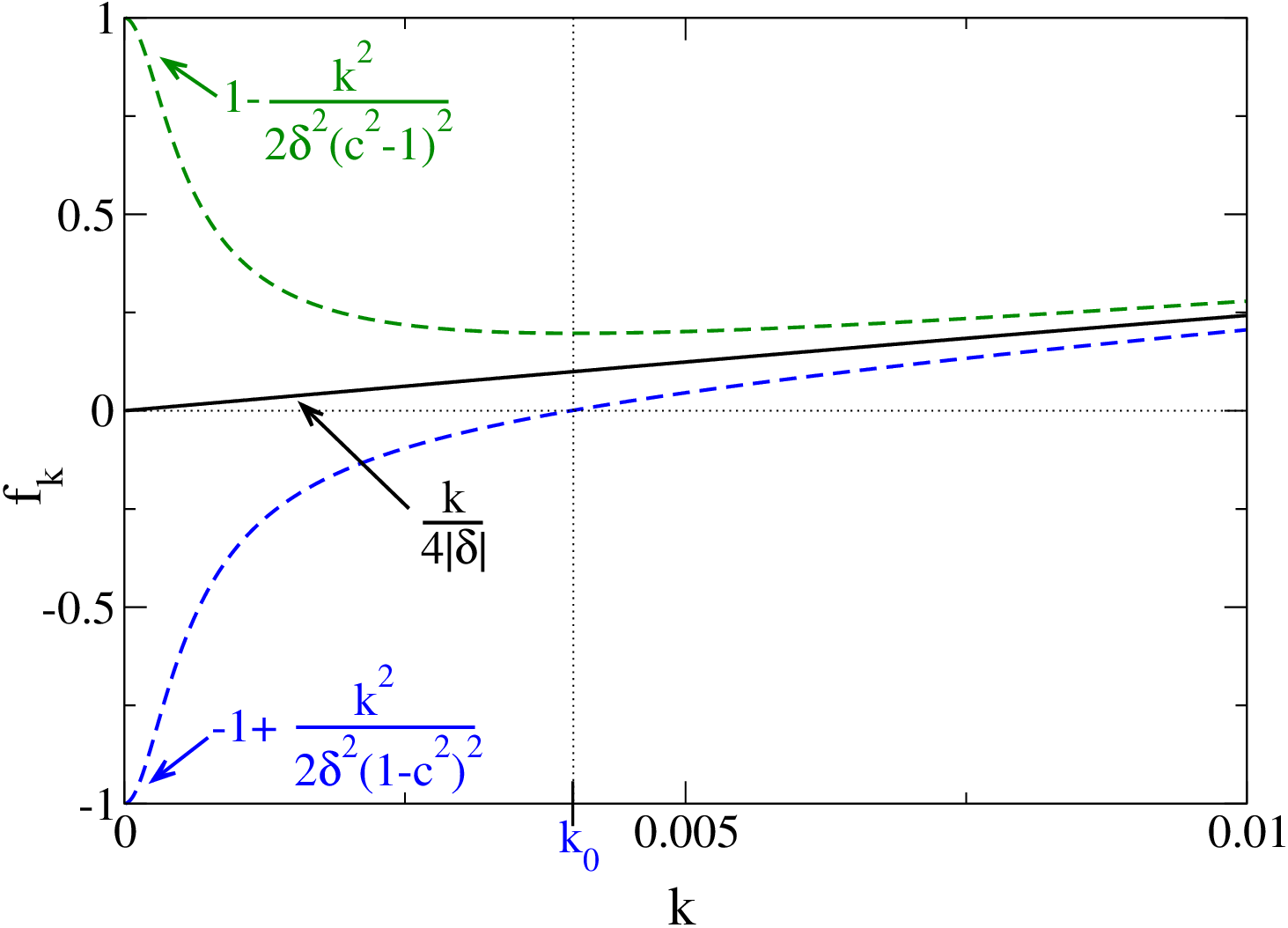}
\caption{(Color online). Discontinuous change of $f_k$ around $c=1$. The curves from bottom to top correspond to 
$c$ equal to $0.98$, $1$ and $1.02$, respectively. The parameter shift $\delta=0.01$.
Lowest order Taylor expansions of $f_k$ around $k_c=0$ are listed (but not plotted) for all three curves.
When $|c|<1$ then $f_k$ crosses zero for $k_0 \neq 0$. 
}
\label{fig19}
\end{figure}

\noindent 
Before we discuss $f_k$ in details, we start the calculation of fidelity 
by approximating the product in (\ref{Fprod_MPS}) by an integral.
Similarly as in Sec. \ref{Sec_XX} we introduce 
$\ln {\cal F}_{\rm smooth} = \frac{N}{2\pi}\int_0^\pi dk \ln|f_k|$. 
In order to calculate it we use approximation (\ref{tildefk_MPS}), 
change the variable $k$ into $|\delta| l$ and send the new upper integration limit, $\pi/|\delta|$, to infinity. This yields
$$
\ln {\cal F}_{\rm smooth} \simeq \frac{N}{4 \pi} |\delta| \int_0^\infty dl \ln \left( \frac{(l^2/4 +  (c^2-1))^2}{(l^2/4 +  (c^2-1))^2+l^2} \right).
$$
The last integral can be calculated analytically:
\begin{equation}
\frac{\ln {\cal F}_{\rm smooth}}{N} = -|\delta| A_{MPS}(c) + {\cal O} (\delta^2),
\label{fsmooth_MPS}
\end{equation}
where $A_{MPS}(c)$ is defined by (\ref{Ac_MPS}). 
We notice that smooth (continuous) representation of fidelity 
reproduces leading exponential decay from (\ref{fidel_MPS3}). 
It also agrees with the general scaling results (\ref{near_F}) and (\ref{away_F}). 
The ${\cal O}(\delta^2)$ error in the integral can be verified numerically 
in a similar way as described in the Appendix. 

As noticed earlier, $A_{MPS}(c)$ is nonanalytic for $|c|=1$. This is also 
visible from momentum space perspective through discontinuity of $f_k$ at $k_c=0$ and $|c|=1$. 
Indeed, it is easy to check that
$$
\begin{array}{lr}
\lim_{k\to k_c} f_k = -1& {\rm for}  ~ |c|<1, \\
\lim_{k\to k_c} f_k = 0 &  {\rm for} ~ |c|=1, \\
\lim_{k\to k_c} f_k = 1 &  {\rm for} ~ |c|>1.
\end{array}
$$
This discontinuity is presented in detail in Fig. \ref{fig19}.

Appearance of oscillations and some constant prefactors in (\ref{fidel_MPS3}),
overlooked by ``continuous"  ${\cal F}_{\rm smooth}$,  can be explained in a similar 
way as in Secs. \ref{Sec_Ising} and \ref{Sec_XX}. Again they can be traced
back to the fact that $f_k$ crosses zero when $|c|\le1$ (Fig. \ref{fig18}).
This leads  to logarithmic 
divergence of $\ln|f_k|$ causing corrections to  ${\cal F}_{\rm smooth}$.

More quantitatively, when $|c|=1$ then $f_k$ reaches zero for $k=k_c=0$. 
Similarly as in Sec. \ref{Sec_Ising}  it leads to the additional term, saturating at $\ln 2/2N$, 
which has to be added to the right-hand side of (\ref{fsmooth_MPS}).  
It results in the prefactor of $\sqrt{2}$ in (\ref{fidel_MPS3}).
 When $|c|<1$, $f_k$ reaches zero for $k_0\neq0$, which is different than $k_c =0$ for which the gap closes. 
It is easy to check that 
\begin{equation}
k_0 = \arccos\left( \frac{1-\delta^2(1-c^2)}{1+\delta^2(1-c^2)}\right) \simeq 2 |\delta| \sqrt{1-c^2}.
\label{k0_MPS}
\end{equation}
This $k_0$ is in general incommensurate with momenta $k$ (\ref{Fprod_k_MPS}) and it leads to oscillations of fidelity. 
We can directly use the formalism presented in Sec. \ref{Sec_XX} to obtain the additional factor of 
$2\left|\cos (k_0 N /2)\right|\approx 2\left|\cos(|\delta|N\sqrt{1-c^2})\right|$ modifying 
${\cal F}_{\rm smooth}$.  Note that this is 
the prefactor appearing in (\ref{fidel_MPS3}) and that the approximation (\ref{k0_MPS}) has been 
used to simplify $\cos (k_0 N/2)$. 


\section{Fidelity and nonequilibrium quantum quenches}
\label{instant_quench}
Our goal here is to briefly illustrate  how our results and analytical techniques 
presented in the previous sections 
can contribute to a better theoretical 
description of quantum quenches (both instantaneous and 
continues ones).

The simplest quench one can consider is the instantaneous quench, where the
parameter of the Hamiltonian, say $\lambda$, is changed at once from $\lambda_1$ to
$\lambda_2$. Assuming that the system was initially prepared in the  ground state
$|\lambda_1\rangle$, the modulus of the overlap of its wave-function 
onto the new ground state $|\lambda_2\rangle$
 reads (\ref{def_F})
\begin{equation}
|\langle \lambda_1|\lambda_2\rangle| = {\cal
F}(\lambda_1/2+\lambda_2/2,\lambda_2/2-\lambda_1/2).
\label{InstF}
\end{equation}
In other words, fidelity provides here a square root of the probability 
of finding the system in the new ground state
after the instantaneous  quench. If the system is in the thermodynamic
limit and $\delta=\lambda_1/2-\lambda_2/2$ is small enough, depending on $\lambda=\lambda_1/2+\lambda_2/2$
either  (\ref{near_F}) or (\ref{away_F}) can be used to evaluate (\ref{InstF})
near the critical point ($\lambda \approx \lambda_c$).

Another application of the analytical techniques developed in the former sections
comes in the context of the density of  quasiparticles excited during an instantaneous 
quench \cite{deGrandi2010a,deGrandi2010b,PolkovnikovArXiv2010}. 
For clarity of
our discussion we will focus here on the instantaneous transition in the Ising model, where 
the system is suddenly moved from $(g_1,\gamma_1)$ to $(g_2,\gamma_2)$ with 
$g_{1,2}=1+\epsilon\pm\delta$ and $\gamma_{1,2}=1$ (path A on Fig. \ref{fig1}; see (\ref{XYH}) for the 
Hamiltonian). 
The density of quasiparticles reads
\begin{equation}
n_{ex} =\int_0^\pi\frac{dk}{\pi} ~ p^{\rm ex}_{k}  = \int_0^\pi \frac{dk}{\pi}(1-f_k^2)= |\delta| B(c)+{\cal O} (\delta^2),
\label{nex_ising}
\end{equation}
where $p_k^{\rm ex}$ is the probability of excitation of the $k$-th momentum mode during the quench 
\cite{PolkovnikovArXiv2010,Deng2011},
$f_k$ is related to fidelity (\ref{InstF}) through (\ref{Fprod}-\ref{Fprod_qk}),
and again we assume $|\delta|$, $|\epsilon|\ll1$ and $c=\epsilon/|\delta|$.

\begin{figure}
\includegraphics[width=\columnwidth,clip=true]{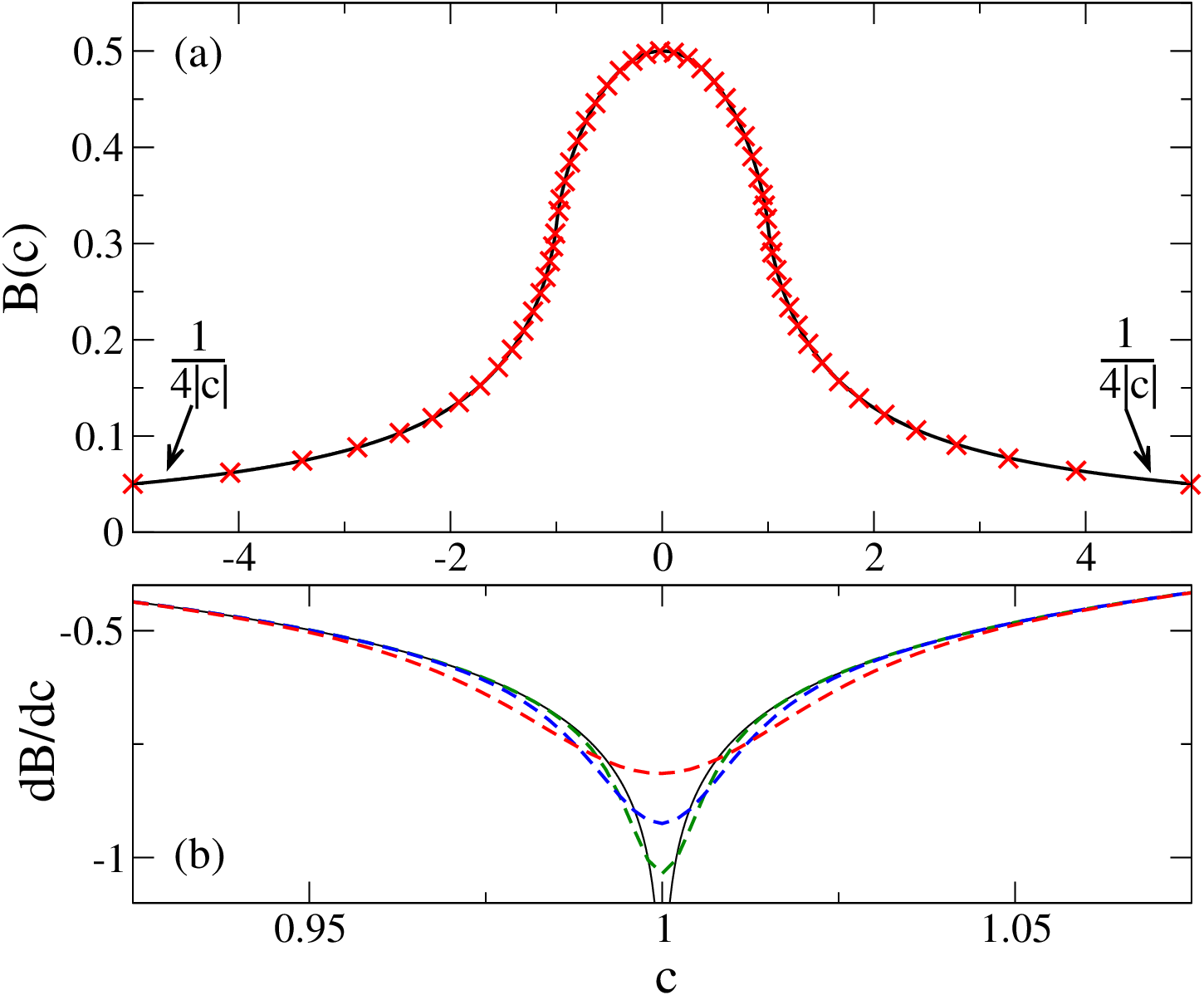}
\caption{(Color online). 
Scaling function (\ref{Bcanal}), and its derivative (\ref{Bderiv}), characterizing 
density of quasiparticles after an instantaneous quench in the Ising model (\ref{nex_ising}). Panel (a): 
the black line shows analytical expression (\ref{Bcanal}), while red crosses show numerically evaluated 
$n_{ex}/|\delta|$ from (\ref{nex_ising})  for $N=10^5$ and $\delta=10^{-3}$.
Panel (b): logarithmic divergence of $dB(c)/dc$ around $c=1$ (\ref{Bderiv}). 
The solid black line shows analytical result (\ref{Bderiv}). 
The dashed lines present  numerics 
done for $\delta=10^{-3}$ and $N$ equal to $10^5$ (red), $2\times10^5$ (blue) and $4\times10^5$ (green) [top to bottom]. 
The logarithmic singularity is reached in the limit of $N\rightarrow \infty$ 
(compare with Fig. \ref{fig4} and discussion in Sec. \ref{Sec_Ising}).
}
\label{fig20}
\end{figure}

\noindent The scaling function $B(c)$  -- presented in Fig. \ref{fig20}a -- is given by
\begin{equation}
B(c)= \frac{1}{2 \pi}\left \{
\begin{array}{ll} 
\begin{split}
& (1-|c|)  {\rm Im} E(c_2) + 2 K(c_1) & {\rm for}~ |c|<1,  \\
& (|c|-1)  {\rm Im} E(c_2) - 2K(c_1) & {\rm for}~ |c|\ge1,
\end{split}
\end{array}
\right. 
\label{Bcanal}
\end{equation}
where $c_1$, $c_2$, $K$ and $E$ are given by (\ref{c1c2}) and (\ref{KEeliptic}).
To obtain this result we used the approximation (\ref{tildefk_Ising}).

\noindent Equation (\ref{nex_ising}) extends the results of \cite{PolkovnikovArXiv2010} 
where density of excited quasiparticles $n_{ex}$ was calculated for a specific situation 
where the instantaneous quench has either began from the critical point or 
ended at the critical point. Note that we consider the instantaneous quench everywhere 
around the critical point.
Moreover, our calculation provides the scaling function $B(c)$, which 
is indispensable for  capturing full universal description of $n_{ex}$ for any two states close to the critical point.
It also allows for unified description of the density of quasiparticles for
instantaneous quenches both near and away from the critical point. To
illustrate the latter, we consider $|c|\gg 1$, but still $|\delta| \ll |\epsilon| \ll 1$, and 
expand (\ref{Bcanal}) getting $B(c) \simeq 1/4|c|$. Putting it into (\ref{nex_ising}) we obtain
$$
n_{ex}  \simeq \delta^2/4|\epsilon|.
$$

\noindent We also note that the scaling function $B(c)$ is nonanalytic for $|c|=1$. 
This results from  discontinuity of $f_k$, and consequently $p_k^{\rm ex}$,
at $k_c=0$ and  $|c|=1$ (\ref{disco}). 
Indeed, the derivative of $B(c)$ can be expanded around $c=1$ 
\begin{equation}
\left.\frac{dB(c)}{dc}\right|_{c\rightarrow1} = \frac{2-3 \ln 2}{2 \pi}+\frac{\ln |1-c|}{2 \pi} + {\cal O}(c-1).
\label{Bderiv}
\end{equation}
This logarithmic divergence is reached when the size of the system $N\rightarrow \infty$ 
(see Fig. \ref{fig20}b and the related  discussion in Sec. \ref{Sec_Ising}).

\begin{figure}[t]
\includegraphics[width=\columnwidth,clip=true]{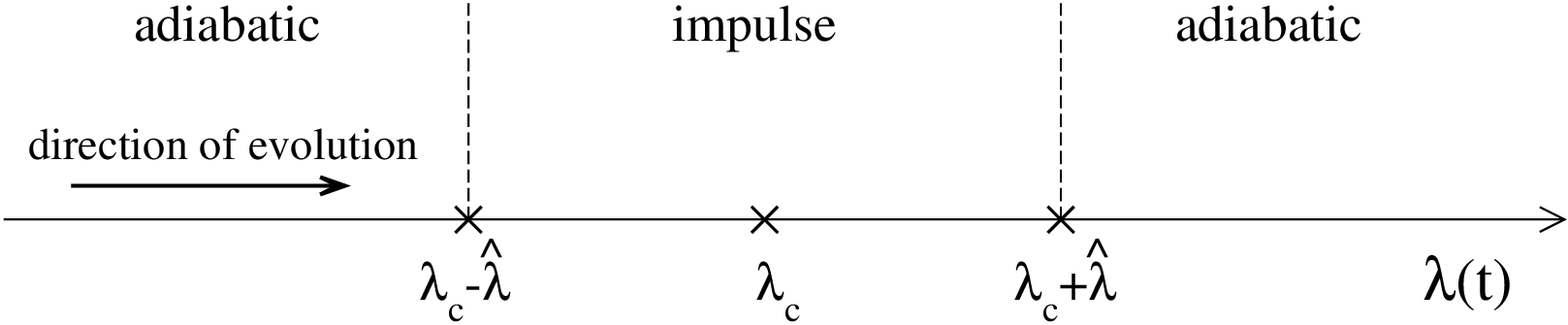}
\caption{Schematic illustration of dynamics of a generic quantum phase transition (see text for details).}
\label{fig21}
\end{figure}

A more complicated quench dynamics shows when the parameter driving the
transition, say $\lambda$,  is gradually shifted 
$$
\lambda(t) = \lambda_c + t/\tau_Q, \ \ t =-\infty \dots \infty,
$$
where the quench time scale $\tau_Q$ controls how fast the system is driven: 
see \cite{DziarmagaReview}
for a review of the resulting quantum dynamics from the perspective relevant for our
discussion. Briefly, it turns out that the system's evolution can be {\it approximately} 
divided into three regimes (Fig. \ref{fig21}) \cite{BD2005,dorner}. First, the adiabatic regime 
takes place as long as the system is far enough from the critical point:  
$\lambda(t)\le\lambda_c-\hat \lambda$, where
$\hat\lambda$ is a small parameter that will be discussed below. Assuming that
the evolution started from a ground state, 
wave-function of the driven system is given here  by the instantaneous ground state 
of its   Hamiltonian. Then the impulse (diabatic) regime happens around the 
critical point,  $\lambda(t)\in(\lambda_c-\hat\lambda,\lambda_c+\hat\lambda)$,
and system's  wave-function is assumed not to change here. Therefore, it is 
the same as at the last ``adiabatic'' instant $\hat t$ such that $\lambda(-\hat t)=\lambda_c-\hat\lambda$.
Finally, the system enters the second adiabatic regime when $\lambda(t)\ge \lambda_c+\hat\lambda$. 
It is assumed that no additional excitations are created by the quench in this
regime. The parameter 
$\hat\lambda$ is provided by the quantum version of the  Kibble-Zurek theory \cite{DziarmagaReview}
$$
\hat\lambda \sim \tau_Q^{-1/(1+z\nu)},
$$
where $\tau_Q\gg1$ is assumed (the slow quench limit) and the prefactor
omitted above is ${\cal O}(1)$: $\hat\lambda\ll1$. In addition, one also 
assumes here that the correlation length ($\xi$)  during the quench is much 
smaller than the system size to avoid finite size effects. 
In the adiabatic-impulse approximation this implies that 
$L\gg\xi(\lambda_c-\hat\lambda) \sim \hat\lambda^{-\nu}$ or simply 
$N/\tau_Q^{d\nu/(1+z\nu)}\gg1$, where $N=L^d$.

It turns out that one can relate all these concepts to fidelity 
and use (\ref{near_F}) to quantify scaling of the system excitation 
with the quench rate $\tau_Q$. Employing the 
adiabatic-impulse approximation, we see  that the population left in the 
instantaneous ground state away from the critical point -- i.e.,  after 
leaving the impulse regime --  is given by 
$$
|\langle \lambda_c-\hat\lambda|\lambda_c+\hat\lambda\rangle|^2 = 
{\cal F}^2(\lambda_c,\delta), \ \ \delta \sim \tau_Q^{-1/(1+z\nu)},
$$
or, after using (\ref{near_F}), by 
\begin{equation}
\exp\left(-N\times {\rm const}/\tau_Q^{d\nu/(1+z\nu)}\right),
\label{KZP}
\end{equation}
where the ${\rm const}$ prefactor is given by ${\cal O}(2A(0))$.
This result has a simple interpretation in the context of symmetry breaking phase transitions.
As predicted by the Kibble-Zurek theory, such transitions lead to creation 
of topological defects whose density is given by $\tau_Q^{-d\nu/(1+z\nu)}$ \cite{DziarmagaReview}.
Thus, the probability of finding the system after the quench in its
instantaneous ground state is exponential in the number of topological 
defects created during the quench (see Sec. V of \cite{Marek_decoh} for the same
conclusion worked out in the Ising model; note that (\ref{KZP}) is
 much smaller than unity  when the adiabatic-impulse approximation applies).

This is in agreement with studies of dynamics of the quantum
Ising chain, whose Hamiltonian is given by (\ref{XYH})
with $\gamma=1$ (note that $g$ in (\ref{XYH}) corresponds to $\lambda$ here) \cite{Marek_decoh,JacekPRL2005}. 
In Sec. V of \cite{Marek_decoh} the Ising chain was initially exposed to a large magnetic field: evolution begun 
from a paramagnetic phase ground state. The magnetic field was ramped
down on a time scale $\tau_Q$ and the system was moved across the
paramagnetic-ferromagnetic critical point. The evolution stopped at the zero
magnetic field (deeply in the second adiabatic regime) and 
the squared
overlap between the evolved wave-function and the ground state at the zero magnetic field was found to equal  
$$
\exp\left(-N\times{\rm const}/\sqrt{\tau_Q}\right),
$$
where ${\rm const}={\cal O}(1)$. This is 
in agreement with (\ref{KZP}) because $\nu=z=d=1$ for the quenched 
Ising model. Naturally, it would be also interesting to 
verify our simple scaling result (\ref{KZP}) in other models as well.

\noindent At the risk of laboring the obvious, we note that fidelity -- 
or more interestingly the scaling factor $\tilde d$ (\ref{tilded}) -- is {\it insensitive} to both 
the relaxation processes after the instantaneous quench and the adiabatic dynamics
following crossing of the critical point. This shall make it a useful tool for theoretical 
characterization of quantum  quenches.


\section{Conclusions}
\label{Sec_con}
We have extensively studied ground state fidelity in the XY model 
illustrating its rich behavior around different critical lines.
This supports its potential  applications as an 
insightful probe of quantum criticality \cite{GuReview}.
A special stress has been placed on discussion of fidelity from the momentum
space angle allowing for better understanding of the studied models.
We have derived a variety of new analytical results in the thermodynamic limit and
verified them through numerical simulations. 

First, we have discussed in detail an example where fidelity is dominated by the 
universal contribution captured by the  scaling relations
(\ref{near_F},\ref{away_F})  derived in \cite{MMR2010}  (Sec. \ref{Sec_Ising}). 
Second, we have focused on the 
case, where close to the critical point the ``smooth'' universal contribution to fidelity 
is strongly modulated: it is multiplied by an oscillating factor (Sec. \ref{Sec_XX}).
This interesting effect has been qualitatively explained and accurately analytically described.
Third, we have characterized fidelity calculated along one of the critical 
lines where non-universal contributions  provide the key input 
(Sec. \ref{Sec_NON}). 
Fourth, we have discussed fidelity near the multicritical point, where
the universal scaling relations are violated and a more advanced scaling theory 
needs to be worked out (Sec. \ref{Sec_Lif}). 

The techniques used in this article may be applicable to other 
systems as well. They have been tested outside of the XY model in the 
extended Ising model (Sec. \ref{Sec_MPS}). They have also been applied to 
get some additional analytical insight into dynamics of quantum phase transitions (Sec.
\ref{instant_quench}). 

From the technical aspects of our work, we would like to stress 
non-triviality of the exchange of the product of over momentum contributions
 into an integral over their logarithms. More fundamentally, we mention
 non-commutativity of $N\to\infty$
and $\delta\to0$ limits near the critical point. Depending on the order in which they  are
taken, one ends in either the ``small system'' limit or the thermodynamic
limit and fidelity behaves very differently in these two limits around the critical 
points. 

\begin{center}
\bf Acknowledgement
\end{center}

This work is supported by U.S. Department of Energy through the LANL/LDRD Program.


\appendix

\renewcommand{\thesection}{}
\section{Estimation of the errors of analytical approximations}
\renewcommand{\theequation}{A\arabic{equation}}

Calculation of the leading universal contribution to fidelity in  Secs. \ref{Sec_Ising} and \ref{Sec_XX}
requires integration of $\ln f_k$ over $k$ from $0$ to $\pi$. Even though $f_k$'s are
known exactly (\ref{Fprod_fk}-\ref{Fprod_qk}), 
such an integral cannot be analytically done to the best of our
knowledge. To proceed, we have approximated $\ln f_k$ in  Secs. \ref{Sec_Ising} and \ref{Sec_XX}
by (\ref{tildefk_Ising}) and (\ref{szlag}), respectively, and extended the integration range over $k$.
Below we estimate the errors resulting from these approximations.

{\bf Approximations in Sec. \ref{Sec_Ising}.} 
We demonstrate here that the difference between our
analytical expression  (\ref{z}) and the exact result scales as ${\cal O}(\delta^2/
\gamma^3)$. Namely, we are going to show that 
\begin{equation}
 E= \frac{1}{2 \pi} \int_0^\pi dk \ln f_k - \left( -|\delta| A(c) / \gamma \right) 
\label{A2_error}
\end{equation} 
is ${\cal O}(\delta^2/\gamma^3)$. The first term on the right-hand side of
(\ref{A2_error}) is the exact expression [$f_k$ is given there by (\ref{Fprod_fk},\ref{pk1},\ref{qk1})],
while the second term is its approximation (\ref{z}). Note that
we do not study here the errors coming from exchange of the product over $f_k$
into the integral (\ref{Fintegral}).

\begin{figure}
\includegraphics[width=\columnwidth,clip=true]{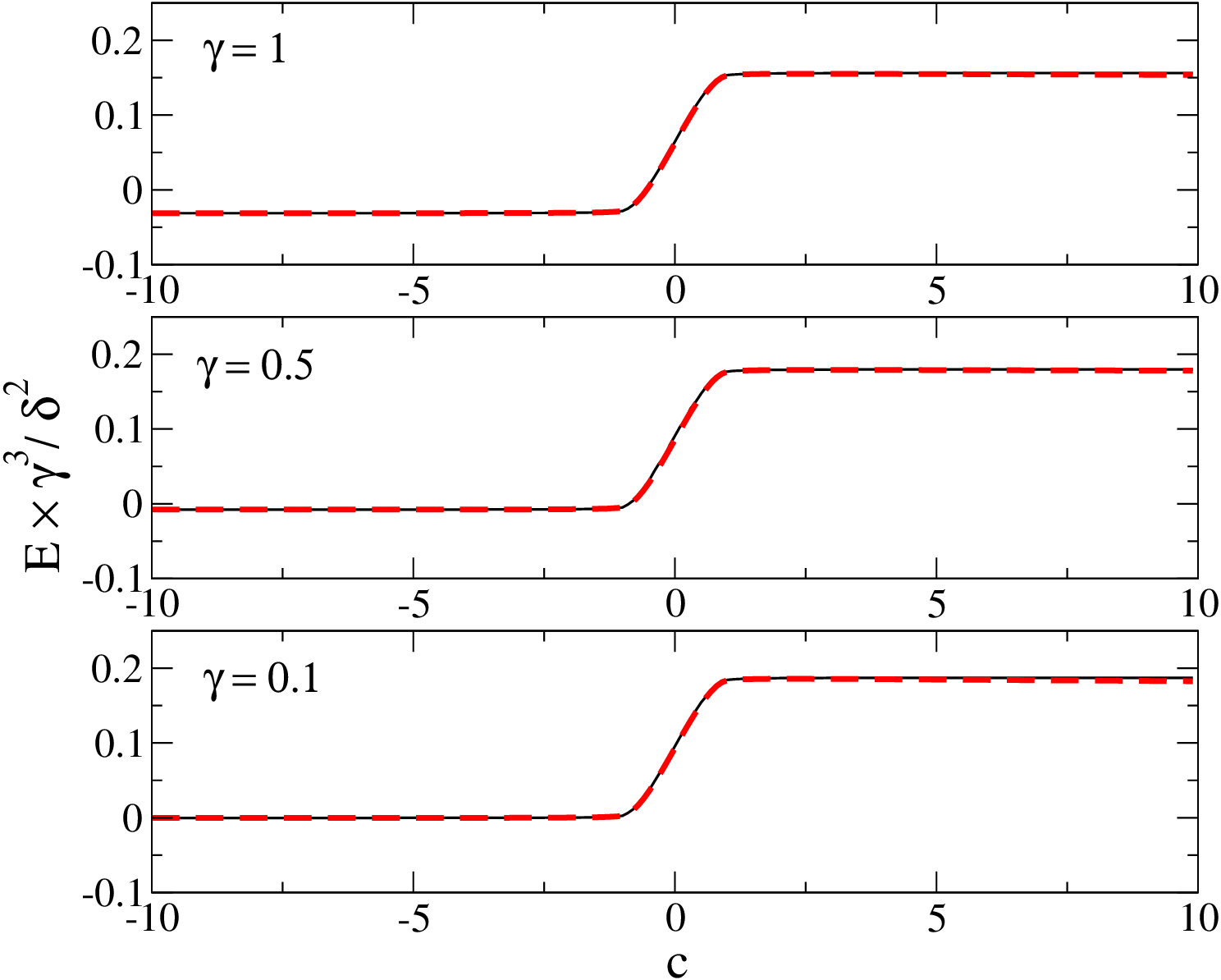}
\caption{(Color online).  
Numerical calculation of the error (\ref{A2_error}) for different 
$\gamma$, $\delta$ and $c=\epsilon/|\delta|$. 
All the curves show that  
$|E|\times\gamma^3/\delta^2$ is bounded from above by about $0.2$. They also illustrate 
that $E$ scales as  $\delta^2/\gamma^3$. 
Upper panel: $\delta=10^{-3}$ (dashed red line) and $\delta=10^{-5}$ (solid black line);  $\gamma=1$
for both curves. Middle panel:   $\delta=10^{-4}$ (dashed red line) and 
$\delta=10^{-6}$ (solid black line);  $\gamma=0.5$ for both curves.
Bottom panel: $\delta=10^{-5}$ (dashed red line)  and $\delta=10^{-7}$ (solid black line); 
$\gamma=0.1$ for both curves. We keep $|\epsilon|, |\delta|\ll\gamma^2$ in
these calculations.
}
\label{fig22}
\end{figure}

For simplicity, we  restrict our analytical studies to the  specific case when one of the states
is exactly at the critical point: $\epsilon=\delta>0$ ($c=1$). 
Straightforward, but tedious,
generalization of the proof listed below allows to extend  it to any $|\epsilon| \ll \gamma^2$.
A numerical check  that the same error appears when $\epsilon\neq\delta$ ($c\neq1$) is 
illustrated in Fig. \ref{fig22}.

\noindent
Error (\ref{A2_error}) is bounded: $|E| <  E_1 +  E_2$, where 
\begin{eqnarray}
E_1 &=&\left|\frac{1}{2 \pi} \int_0^\pi dk \ln f_k -\frac{1}{2 \pi} \int_0^\pi dk \ln \tilde f_k \right| 
\nonumber \\ &\le&
\frac{1}{2 \pi} \int_0^\pi dk \left|\ln f_k -\ln \tilde f_k\right| 
\label{A2_error1}
\end{eqnarray} 
is due to approximation of $p_k$ and $q_k$ in (\ref{tildefk_Ising}) by $\tilde
p_k$ and $\tilde q_k$ (see below) and
$$
E_2 = \left| \frac{1}{2 \pi} \int_\pi^\infty dk \ln \tilde f_k \right|
\le \frac{1}{2 \pi} \int_\pi^\infty dk \left|\ln \tilde f_k \right|
$$
results from sending the upper integration limit in (\ref{fid_ising_app}) to infinity.
For clarity of our discussion, we repeat the expressions for exact $f_k$
\begin{eqnarray*}
\ln f_k &=& \frac12 \ln \left(\frac12 + \frac{p_k}{2\sqrt{p_k^2+q_k^2}}
\right),  \\
p_k &=& (1-\cos k) (1-\cos k +2 \delta) + \gamma^2 \sin^2 k,  \\
q_k &=& 2 \delta \gamma \sin k, 
\end{eqnarray*}
and approximated $\tilde f_k$
\begin{eqnarray*}
\ln \tilde f_k &=& \frac12 \ln \left(\frac12 + \frac{\tilde p_k}{2\sqrt{\tilde
p_k^2+\tilde q_k^2}}  \right),  \\
\tilde p_k &=& \gamma^2 k^2, \quad \tilde q_k = 2 \delta \gamma k. 
\end{eqnarray*}

\noindent In order to estimate $E_1$   we find a function $g_k$ such that
\begin{equation}
\left |\ln f_k - \ln \tilde f_k \right| < g_k.
\label{B1}
\end{equation}

To further simplify our calculations, we assume that $\delta/\gamma^2<0.2$: other bounds
 would give us a slightly different prefactor in the estimation of the magnitude of $E$, but will not
 change its scaling with $\gamma$ and $\delta$.
To derive (\ref{B1}) we start with inequality: 
\begin{equation}
|\ln f_k - \ln \tilde f_k| <  0.6\left|f_k^2/\tilde f_k^2 -1 \right|, 
\label{B4}
\end{equation}
which is true when $\left|f_k^2/\tilde f_k^2-1\right| < 0.25$ (the latter will be proved below). We consider 
\begin{equation}
\left|f_k^2/\tilde f_k^2-1\right| = \left|
\frac{\frac{p_k}{\sqrt{p_k^2+q_k^2}} - \frac{\tilde p_k}{\sqrt{\tilde
p_k^2+\tilde
q_k^2}}}{1+
\frac{\tilde p_k}{\sqrt{\tilde p_k^2+\tilde q_k^2}}} \right|.
\label{app2_loge}
\end{equation}
Two remarks are in order now. First, we will use below that 
$p_k,q_k,\tilde p_k,\tilde q_k \ge 0$ and $0\le p_k / q_k\le\infty$.
Second, we will consider two regions in $k$
separately:

$\bullet$ $k \in [0, 2\delta/\gamma]$ or equivalently  $ 0 \le \tilde p_k /
\tilde q_k <1$. \\In this case we estimate (\ref{app2_loge}) as
\begin{eqnarray*}
\left|f_k^2/\tilde f_k^2-1\right|&=&\frac{\left|\frac{p_k/q_k}{\sqrt{1+p_k^2/q_k^2}} -
\frac{\tilde
p_k/\tilde q_k}{\sqrt{1+\tilde p_k^2/\tilde q_k^2}} \right|}{1+\frac{\tilde
p_k/\tilde q_k}{\sqrt{1+\tilde p_k^2/\tilde q_k^2}}}  \\
&\le & \left|\frac{p_k/q_k}{\sqrt{1+p_k^2/q_k^2}} - \frac{\tilde p_k/\tilde
q_k}{\sqrt{1+\tilde p_k^2/\tilde q_k^2}} \right|  \\
&\le&\left| p_k/q_k -  \tilde p_k/\tilde q_k \right| = g_k / 0.6. 
\end{eqnarray*}
Above, in the first step we bounded denominator by $1$ and in the second step we
used inequality $\left|x/\sqrt{1+x^2}-y/\sqrt{1+y^2}\right| \le
\left|x-y \right|$. Subsequent study of $|p_k/q_k -  \tilde p_k/\tilde q_k|$ 
reveals that $\left|f_k^2/\tilde f_k^2 -1\right| < 0.25$ 
when $\delta/\gamma^2 < 0.2$. This in turn validates inequality (\ref{B4}) in the considered $k$-region. 
The above estimation provides us also with the  expression for
$g_k$ from (\ref{B1}): $g_k=0.6|p_k/q_k -\tilde p_k/\tilde q_k|$.
We integrate it over $k$, and simplify the result getting
\begin{eqnarray}
\label{B7}
\frac{1}{2 \pi}\int_0^{2 \delta/\gamma} dk \, g_k < 0.11~ \delta^2 /\gamma^3.
\end{eqnarray}

$\bullet$ $k \in [2 \delta/\gamma, \pi]$ or equivalently  $ 0 \le \tilde q_k / \tilde p_k <1$. \\
In that case we estimate (\ref{app2_loge}) as
\begin{eqnarray*}
\left|f_k^2/\tilde f_k^2-1\right|&=&\frac{\left|\frac{1}{\sqrt{1+q_k^2/p_k^2}} -
\frac{1}{\sqrt{1+\tilde q_k^2/\tilde p_k^2}}
\right|}{1+\frac{1}{\sqrt{1+\tilde
q_k^2/\tilde p_k^2}}}  \\
&\le&\frac{1}{1+1/\sqrt{2}}\left|\frac{1}{\sqrt{1+q_k^2/p_k^2}} -
\frac{1}{\sqrt{1+\tilde q_k^2/\tilde p_k^2}} \right|   \\
&\le&\frac{1}{2+\sqrt{2}}\left| q_k^2/p_k^2 -  \tilde q_k^2/\tilde p_k^2
\right|=g_k/0.6, 
\end{eqnarray*}
where in the first step  we bounded denominator by $1+1/\sqrt{2}$ and later we used
inequality $\left|1/\sqrt{1+x^2} - 1/\sqrt{1+y^2} \right| \le \left| x^2 - y^2 \right|/2$.
A careful study of $\frac{1}{2+\sqrt{2}}\left| q_k^2/p_k^2 -  \tilde q_k^2/\tilde p_k^2\right|$
shows that $\left|f_k^2/\tilde f_k^2 -1\right| < 0.25$ when $\delta/\gamma^2 < 0.2$. It validates 
inequality (\ref{B4}) in the studied $k$-region. From the above calculation we see that 
 $g_k$  from (\ref{B1}) is equal to $\frac{0.6}{2+\sqrt{2}}\left| q_k^2/p_k^2 -  \tilde q_k^2/\tilde p_k^2\right|$.
 It can be integrated over $k$ and the obtained result can be simplified to yield
\begin{equation}
\frac{1}{2\pi} \int_{2\delta/\gamma}^\pi dk ~g_k < 0.25 ~ \delta^2/\gamma^3.
\label{B10}
\end{equation}
This concludes our proof of (\ref{B4}) for $\delta/\gamma^2<0.2$.

Combining (\ref{A2_error1}) with (\ref{B1}), (\ref{B7}) and (\ref{B10}) we obtain
\begin{equation}
E_1 < 0.36 ~ \delta^2/\gamma^3.
\label{qwerty1}
\end{equation}

\begin{figure}[t]
\includegraphics[width=\columnwidth,clip=true]{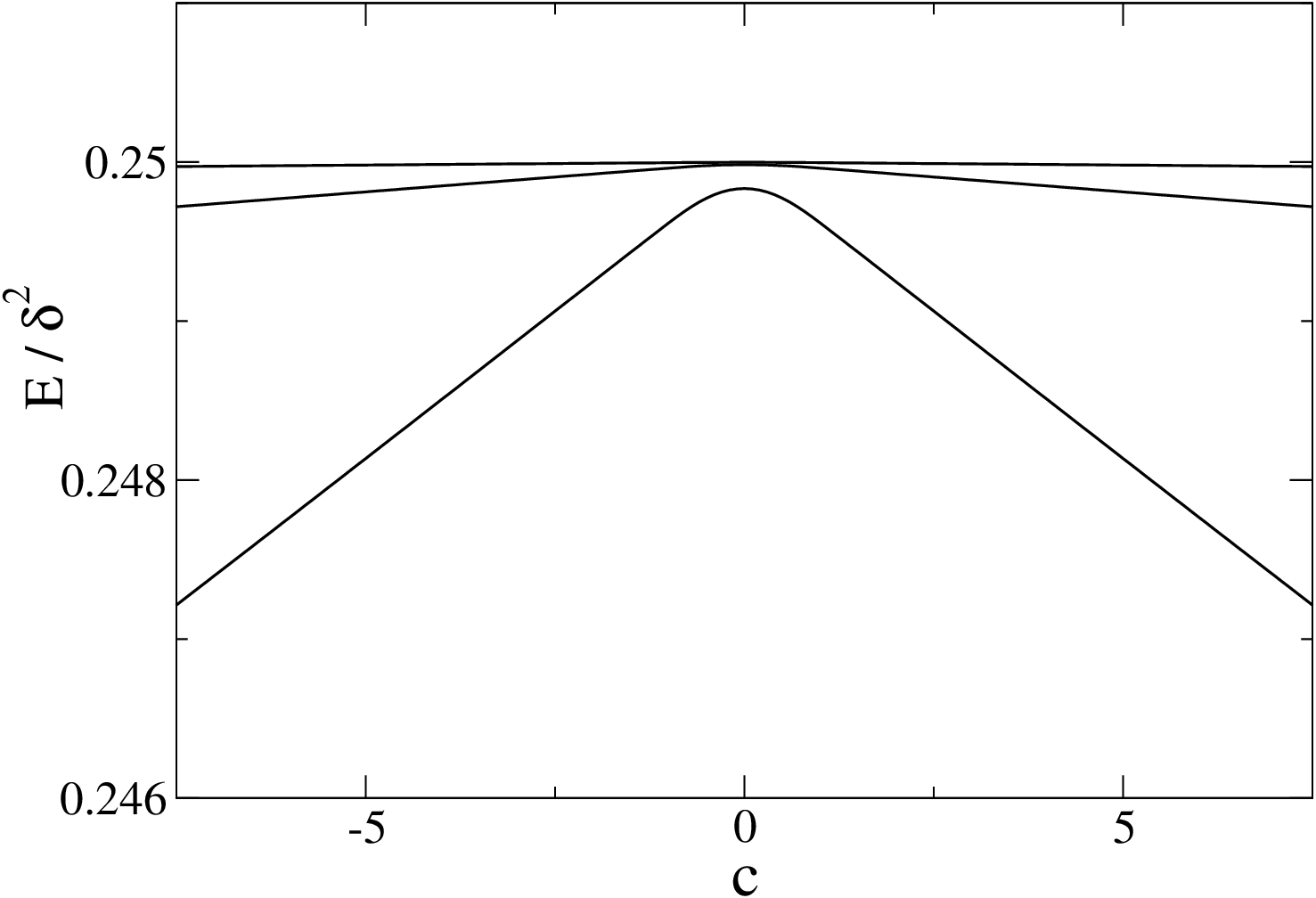}
\caption{Integration error (\ref{nadejszla}).
The curves from bottom to top correspond to $\delta = 10^{-3}$, $10^{-4}$ and $10^{-5}$, respectively.
For each $\delta$ we choose $g=0, 0.9, 1$. The curves for different $g$'s, but the same 
$\delta$, perfectly collapse onto each other. This plot supports
$E\approx 0.25\,\delta^2$. 
}
\label{fig23}
\end{figure}

Estimation of $E_2$ is straightforward because 
\begin{eqnarray}
\label{qwerty}
&&E_2\le\frac{1}{2 \pi} \int_{\pi}^\infty dk | \ln \tilde f_k |=  \\
&&\frac{|\delta|}{4 \pi \gamma} \int_{\pi \gamma /\delta }^\infty dl\left| \ln
\left( \frac{1}{2}+\frac{1}{2\sqrt{1+4/ l^2}}  \right) \right|< \nonumber \\
&&\frac{|\delta|}{4 \pi \gamma} \int_{\pi \gamma /\delta }^\infty\frac{dl}{l^2} =
\frac{\delta^2} {4 \pi^2 \gamma^2} < 0.03 ~ \delta^2 / \gamma^3.\nonumber
\end{eqnarray}

Summing up the contributions (\ref{qwerty1}) and  (\ref{qwerty}) we find  that 
$$
|E| < 0.4\,\delta^2/\gamma^3
$$  
when $\epsilon = \delta$ ($c=1$). This is in agreement 
with Fig. \ref{fig22}, which in fact suggests a stronger bound, and 
extends our  analytical results beyond the $\epsilon=\delta$ case.

{\bf Approximations in Sec. \ref{Sec_XX}.} We justify here numerically 
the ${\cal O} (\delta^2)$ error in 
(\ref{fid_isingXX}). To that end we look at
\begin{equation}
E = \frac{1}{2 \pi} \int_0^\pi dk \ln f_k - \left( -2 |\delta| A(c) \right),
\label{nadejszla}
\end{equation}
where this time $f_k$ is given exactly by (\ref{Fprod_fk},\ref{pk2},\ref{qk2}). 
The numerical results for various values of parameters are presented in Fig. \ref{fig23}.
It shows that the error scales as ${\cal O}(\delta^2)$ confirming that our approximation 
correctly captures leading universal contribution to ${\cal F}_{\rm smooth}$.
We note also that our numerics suggests that the error, 
and in principle the exact value of the integral (\ref{fid_isingXX}),
do not depend on $g \in [-1,1]$.


\begin{thebibliography}{46}
\bibitem[{Fit()}]{Fit}
\bibinfo{note}{We perform least squares linear fitting. The errors that we
  provide correspond to one standard deviation.}
\end{thebibliography}

\end{document}